\documentclass[12pt]{article}

\newcommand{\be}{\begin{equation}} 
\newcommand{\ee}{\end{equation}}
\newcommand{\bea}{\begin{eqnarray}} 
\newcommand{\eea}{\end{eqnarray}}

\usepackage{graphicx,amssymb}

\def\IR{\relax{\rm I\kern-.18em R}}
\def\IN{\relax{\rm I\kern-.18em N}}

\def\vf{\varphi}
\def\C{{\cal C}}
\def\l{\lambda}
\def\z{\zeta}
\def\vf{\varphi}

\font\cmss=cmss10 \font\cmsss=cmss10 at 7pt
\def\IZ{\relax\ifmmode\mathchoice
{\hbox{\cmss Z\kern-.4em Z}}{\hbox{\cmss Z\kern-.4em Z}}
{\lower.9pt\hbox{\cmsss Z\kern-.4em Z}}
{\lower1.2pt\hbox{\cmsss Z\kern-.4em Z}}\else{\cmss Z\kern-.4em Z}\fi}

\begin{document}
\title{Entropic C theorems\\in free and interacting two-dimensional 
field theories} 
\author{J. Gaite\\ 
{\it Instituto de Matem{\'a}ticas y F{\'\i}sica Fundamental, CSIC,}\\ 
       {\it Serrano 123, 28006 Madrid, Spain}\\
and\\
{\it Laboratorio de Astrof{\'\i}sica Espacial y F{\'\i}sica Fundamental,}\\
{\it Apartado 50727, 28080 Madrid, Spain}
}
\date{July 5, 1999}

\maketitle

\begin{abstract}
  The relative entropy in two-dimensional field theory is studied on a
  cylinder geometry, interpreted as finite-temperature field theory.
  The width of the cylinder provides an infrared scale that allows us to
  define a dimensionless relative entropy analogous to Zamolodchikov's
  $c$ function.  The one-dimensional quantum thermodynamic entropy
  gives rise to another monotonic dimensionless quantity.  I
  illustrate these monotonicity theorems with examples ranging from
  free field theories to interacting models soluble with the
  thermodynamic Bethe ansatz.  Both dimensionless entropies are
  explicitly shown to be monotonic in the examples that we analyze.
\end{abstract}
\global\parskip 3pt
{\small
}
\vskip .5cm

\section{Introduction}

It has been shown that the irreversible character of the
renormalization group (RG) can be cast in a sort of $H$ theorem
analogous to Boltzmann's, thus generalizing this theorem from ordinary
time evolution to the evolution with the RG parameter \cite{I-OC}. The
irreversible quantity, the field theory entropy relative to a fixed
point of the RG, is a monotonic function of the coupling constants and
increases in the crossover from one fixed point to another less
stable. However, the Wilson RG picture considered in \cite{I-OC},
wherein one has to deal with all the couplings generated by the RG
action, relevant and irrelevant alike, turns out to be too complex and
was indeed assimilated to a non-equilibrium thermodynamics setting.
One can start with only the relevant couplings but then one must
utilize a different RG which changes some infrared (IR) scale.  A
possibility is to define the field theory on a finite geometry
characterized by some parameter, loosely associated to its size, which
plays the r\^ole of IR scale.  Then the monotonicity theorem for the
relative entropy can be cast as a RG theorem similar to the celebrated
Zamolodchikov $c$ theorem \cite{I}.

Among the various geometries we could consider, the cylinder stands
out for its simplicity. It is defined by only one scale, the length of
the compact dimension, and the finite-size corrections to the
partition function turn out to be computable.  Moreover, on a cylinder
of circumference $\beta$ the monotonicity theorem adopts a form with a
thermodynamic interpretation, the temperature being $T= 1/\beta$
\cite{I}.  Thus the inverse temperature is used as RG parameter,
providing a thermodynamic interpretation of the RG, as in
Ref.~\cite{CaNFr}.  The connection with concepts of $1+1$ QFT at
finite temperature is intellectually appealing and useful for
computational purposes.  For example, finite-size corrections are
calculated in terms of the properties of one-dimensional ($1d$)
quantum gases.  In addition to the relative entropy, the $1d$ quantum
entropy provides another monotonic quantity with a different
interpretation.  We must remark that the definition and monotonicity
of the relative entropy, as exposed in Ref.~\cite{I-OC}, already have a
thermodynamical motivation in the $2d$ context, independently of the
type of geometry.  However, in field theory we prefer to dissociate
the coupling from a thermal interpretation and we reserve the concept
of temperature for its r\^ole in the $1d$ quantum picture.
Nevertheless, it shall be evident that the proofs of the monotonicity
theorems for the $2d$ relative entropy or for the $1d$ quantum entropy
are essentially the same.

In Ref.~\cite{I} these ideas were illustrated only with free field
models and the calculations of the corresponding finite-size
corrections were presented very concisely. We shall begin here with a
more detailed analysis of the properties of both types of entropy, in
particular, considering whether they are universal quantities.  Next,
we proceed to the explicit calculation of the finite-size corrections
for soluble models corresponding to free-field theories, including
thermodynamic quantities as well as the expectation values of the
stress tensor, and hence of the entropic monotonic quantities.  The
properties of these quantities will be displayed in the corresponding
plottings.  Further to free-field models, it will be demonstrated that
interacting models are also suitable for calculation of their
finite-size corrections and monotonic quantities with powerful
methods. In particular, integrable models on the cylinder are
appropriate for application of the thermodynamic Bethe ansatz (TBA).
Plots of the monotonic quantities obtained with this method display
similar behaviour to those of free-field models.

The paper is divided in three parts. The first part is devoted to
formulating the monotonicity theorems for $2d$ field theory and to
giving its thermodynamic interpretation on the cylinder. The second
part applies these theorems to the relatively simple cases of the
Gaussian and Ising models.  They allow for an explicit calculation of
thermodynamic quantities and their connection with the components of
the stress tensor. Section 3 is devoted to interacting models which
lend themselves to computation of thermodynamic quantities. The
essential tool is the thermodynamic Bethe ansatz, which is first
applied to models with purely statistical interaction, resulting again
in explicit expressions for the relevant quantities, and in second
place to models in which the TBA equations have to be solved
numerically. Afterwards there comes a discussion of the results
obtained and, finally, two appendices, the first one on the method for
the computation of finite-size corrections based on the
Euler-MacLaurin formula and the second one on the computation of the
expectation value of the stress tensor on the cylinder for free
theories.

\section{Entropic $C$ theorems}
 
\subsection{General properties of the relative entropy in two-dimensional 
  field theory}

Let us briefly recall some concepts already introduced in
Ref.~\cite{I-OC}.  The field theory probability distribution 
associated to some statistical system is given by
\be 
{\cal P}[\phi,\{\l\}]
= e^{-I[\phi,\{\l\}]+W[\{\l\}]}, 
\ee 
and depends on some stochastic field $\phi$ and a set of coupling constants 
$\{\l\}$.  The quantity $W[\{\l\}]$ is needed for normalization and is
of course minus the logarithm of the partition function.  A composite
field is defined as the derivative of the action with respect to some
coupling constant,
\be f_\l =
{\partial I\over\partial\l}.  \label{cofi} 
\ee 
For example, if we consider the thermal coupling, the coupling constant is 
the inverse temperature and the composite field represents the energy. 
As is usual, we assume for simplicity that $I[\phi,\{\l\}]$ is linear
in the coupling constants.  

The relative entropy, a concept
borrowed from probability theory, turns out to be the Legendre
transform of $W(\l)-W(0)$ with respect to $\l$ \cite{I-OC}:
\be 
S_{\rm rel}(\l) = W(\l) - W(0) - \l{d W\over d\l}
= W - W_0 - \l\,\langle f_\l \rangle. \label{Legtrans} 
\ee 
Obviously, $S_{\rm rel}(0)=0$. Furthermore, as a straightforward
consequence of its definition,
\bea 
\l{dS_{\rm rel}\over d\l} &=& \l{dW\over d\l} -
\l{d\over d\l}\left(\l{dW\over d\l}\right) \nonumber\\
&=&- \l^2 {d^2W\over d\l^2} = - \l^2{d\over d\l}\langle f_\l\rangle =
\l^2\langle (f_\l-\langle f_\l\rangle)^2\rangle \geq 0.
\label{pos}
\eea
For the thermal coupling, $S_{\rm rel}$ has indeed the interpretation
of a real thermodynamic entropy which increases with temperature.  In
other cases, it may or may not have a thermodynamic interpretation but
its properties hold nonetheless.

Some qualifications are in order. In field theory we deal with local
fields, so $f_\l = \int\Phi_\l$, where $\Phi_\l(z)$ is a local
composite field, function of the $2d$ coordinates $z = x_1+i\,x_2$.
We must remark that, although these fields are usually constructed as
actual composites of the basic field $\phi$, the existence of this
field needs not be assumed, as in some modern formulations where it is
replaced by the {\em action principle} \cite{GuiMag}. This remark is
important when we start from a $2d$ conformal field theory.  To
prevent the appearance of ultraviolet (UV) divergences it is
convenient to define all the quantities with a UV cutoff
$\Lambda$ (for example, $W[\l,\Lambda]$) which must be eventually
removed to define universal quantities. Even though $W$ is non
universal, we expect $S_{\rm rel}$ to be \cite{I-OC}. In order to have
universality, we consider RG relevant or marginal couplings: In two
dimensions the scaling dimension of the field $\Phi$ must be such that
$0 \leq d_{\Phi} \leq 2$.  This condition may not be sufficient and
shall be made more precise shortly. $W$ and $S_{\rm rel}$ are
extensive and it is convenient to define the associated specific
quantities dividing by the total volume---or area in two dimensions.
Henceforth, we use specific quantities but keep the same notation for
simplicity.  We are interested in an entropy relative to a RG fixed
point, so we must substract from the coupling constants their values
at that point.  (The fixed-point coupling constants may be null in
some cases.) Finally, there is an assumption of positivity of the
probability distribution implied in the inequality (\ref{pos}), like
in Zamolodchikov's theorem.

To derive a universal expression for the specific $S_{\rm rel}$ we must
analyse its dependence on the UV cutoff. We can use the scaling form
of the specific $W$,
\be
W(\l,\Lambda)= \Lambda^{2}\,
{\cal F}\left({\l^{2/y}\over\Lambda^{2}}\right), \label{scal}
\ee
where $y = 2 - d_{\Phi} > 0$ is the dimension of the coupling $\l$.
For the thermal field, the local energy density, $y$ is the inverse of
the critical exponent $\nu$.  If the scaling function is continuously
differentiable around zero (class $C^1$), and we denote $F_0={\cal F}(0)$, 
$F_1 = {\cal F}'(0)$, $W$ can be expanded as
\be
W(\l,\Lambda)= \Lambda^{2}\,F_0 + F_1\,\l^{2/y} + 
\Lambda^{2}\,o(\Lambda^{-2}),
\label{WF}
\ee
with $o(\Lambda^{-2})$ asymptotically smaller than $\Lambda^{-2}$, hence
resulting in a vanishing term as $\Lambda \rightarrow\infty$.  Given that
the UV divergent term of this expansion cancels in $W(\l)-W(0)$, 
the relative entropy yields a finite result in the infinite cutoff
limit, namely,
\begin{equation}
S_{\rm rel}(\l) = W(\l) - W(0) - \l{dW\over d\l} = 
F_1\,\frac{y-2}{y}\,\l^{2/y}.  
\label{SrelF}
\end{equation}
Thus the significance of the assumed regularity condition on the
scaling function is that it is sufficient to endow the monotonicity
theorem with universality.  One can certainly think of simple
functions that are not class $C^1$.  For example, the function ${\cal
  F}(x) = F_0 - x\,\ln x + o(x)$, which will appear in some of the
models studied later.  

We now examine the question of universality in terms of local
fields. This method will lead us to a more concrete formulation.
Let us begin by writing the monotonicity theorem (\ref{pos}) as
\be 
\l{\partial S_{\rm rel}\over \partial \l} = \l^2 \int \!d^2z\,
\langle:\!\Phi(z)\!:\,:\!\Phi(0)\!:\rangle \geq 0,
\ee
with the use of the definition of normal-ordered composite fields,
$:\!\Phi\!: \: = \Phi - \langle\Phi \rangle$.  We can study the UV
convergence of this integral. As a prerequisite, note that possible UV
divergences in the definition of the composite field $\Phi$ are
removed by the substraction of $\langle\Phi \rangle$. The most
singular part of the correlation function for short distance is given
by \be \langle :\!\Phi(z)\!:\, :\!\Phi(0)\!: \rangle \sim |z|^{-2\,
  d_\Phi}.  \ee Hence, the integral converges if $0\leq d_\Phi <1$,
that is, $1<y\leq 2$. Then the derivative of the relative entropy,
$dS_{\rm rel}/d\l$, is a universal quantity and so is $S_{\rm rel}$,
because the integration constant is fixed by the condition $S_{\rm
  rel}(0) = 0$.  For dimensional reasons, it must adopt a form like
that in (\ref{SrelF}):
\be 
S_{\rm rel}(\l) = B\,\l^{2/y},
\label{Srscal}
\ee 
where $B$ is a constant. In fact, upon inversion of the Legendre
transform this form implies that $W$ has the previous first-order
expansion (\ref{WF}), except in the case of $y=2$.  

Fields $\Phi$ satisfying $0\leq d_\Phi <1$ are called strongly
relevant \cite{KlasMel,ConFlu}. They include the thermal coupling of
the unitary minimal models of conformal field theory (CFT), except the
Ising model, wherein the local energy density has $d_\Phi=1$. 
We shall find that the relative entropy of the Ising
model is indeed non universal.  In principle, fields with $1 \leq
d_\Phi \leq 2$ give rise to a non-universal relative entropy, $S_{\rm
  rel}(\l,\Lambda)$. It is monotonic and essentially independent of
$\Lambda$ as long as $\l^{2/y} \ll \Lambda$, which is the condition
necessary for the continuum field theory of the statistical system to
be meaningful. In this sense, one may consider this non-universal
relative entropy within the philosophy of {\em effective field
  theories}, a term which refers to theories that are not
renormalizable but suitable for calculation of many physical
quantities for scales much lower than the cutoff.

We must remark that the simple power-law forms of the relative entropy
(\ref{Srscal}) and the monotonicity theorem are not very informative,
in the sense that, once we know that $S_{\rm rel}$ is finite, they
follow from dimensional analysis. We thus see the necessity of
introducing a new parameter, for example, through a finite geometry.
We will indeed obtain a richer and more illuminating version of the
relative entropy and the monotonicity theorem when we introduce a
finite geometry.

Let us introduce the stress tensor trace, $\Theta := T_a^a$.  Since
$\Theta$ gives the response to a change of scale and the only scale is
in the coupling constant, it is in general proportional to the
relevant field $\Phi$:
$$\Theta = y\,\l\,\Phi.$$
Hence, we can put the monotonicity theorem
for the specific relative entropy in an interesting form:
\be
\l{\partial S_{\rm rel}\over \partial \l}(\l,\Lambda) = 
{1\over y^2}\int \!\!d^2z\,
\langle:\!\Theta(z)\!:\,:\!\Theta(0)\!:\rangle \geq 0. 
\label{monoSrel}
\ee
We will have the occasion to comment on this form in what follows.

\subsection{Finite-size corrections. The cylinder and one-dimensional 
  thermodynamics.}

So far, the relative entropy has been proved to be monotonic with
respect to the coupling constants.  Now we would like to reformulate
the monotonicity theorem for the relative entropy as showing
irreversibility under the RG. We need to substitute the coupling
constant $\l$ by some quantity which can be interpreted as a RG
parameter.  A common way to introduce a RG parameter is through some
IR scale. For example, we may consider a finite size system with a
characteristic length, such as a strip or cylinder of width $L$.
According to finite-size scaling ideas, the free energy can be split
into a {\em bulk part} and a {\em universal finite-size correction}.
Tha latter constitutes a suitable function to derive a non-trivial
relative entropy. Moreover, one can take advantage of the fact that
the classical partition function on a cylinder of width $\beta$ is
equivalent to the one-dimensional quantum partition function at
temperature $T=1/\beta$ to give the RG a thermodynamic interpretation
\cite{CaNFr}.  Indeed, relevant thermodynamic functions of this
quantum system are given by derivatives with respect to $\beta$.  The
first one is the energy, which has one part independent of the
temperature and another that vanishes at $T=0$, corresponding to the
bulk part and the finite size correction, respectively. The part
independent of the temperature, which is non universal, represents the
ground-state energy. A more interesting quantity is the specific
one-dimensional quantum entropy, which turns out to be universal and
will prove to be the right quantity for a thermodynamic monotonicity
theorem.

Let us then consider the system on a cylinder, equivalent to finite
temperature field theory. The partition function is $Z = {\rm
  Tr}\,e^{-\beta\,H}$, which can be represented as a functional
integral on $S^1 \times \IR$ with $\beta=1/T$ the length of the
compact dimension.  We assume that the specific logarithm of the
partition function on a cylinder of width $\beta$ and length $L$ as $L
\to \infty$ can be split into a bulk part and a finite-size
correction,\footnote{This formula has already been proposed for generic
  dimension $d$, on the grounds of dimensional analysis \cite{CaNFr}.}
\begin{equation}
{-\ln Z\over L} = \beta\,{F\over L} = 
e_0(\Lambda,\l)\, \beta + {C(\beta,\l)\over\beta},   \label{fse}
\end{equation}
where $C(\beta,\l)$ is a universal dimensionless function having a
finite limit as $\beta \rightarrow \infty$. Hence, defining $x=
\beta\,\l^{1/y}$ we write $C(\beta,\l)$ as a single-variable
function, $C(x)$. At a RG fixed point it is proportional to the CFT
central charge, $C(0) =-\pi\,c/6$ \cite{BCN,Aff}.

One can readily calculate the $1d$ energy
\begin{equation}
{E\over L} = -{\partial\ln Z/L\over\partial\beta} = e_0 + 
{1\over\beta^2}\left(
\beta{\partial C\over\partial\beta}-C\right) = e_0  - 
{1\over\beta^2}\left(C -
x{dC\over dx}\right).
\label{E}
\end{equation}
At zero temperature ($\beta \rightarrow \infty$) the system is 
on its ground state and therefore $e_0$ represents the specific 
ground state energy whereas the $C$ part is a finite-size effect. 
{}From the energy, Eq.~(\ref{E}), we can compute the thermodynamic entropy
\begin{equation}
{S\over L} = \beta\,{E-F\over L}  = -{2\over\beta}\left(C -
{x\over 2}\,{dC\over dx}\right),
\label{SwC}
\end{equation}
which is universal, since it contains no contribution from $e_0$.
Moreover, the entropy vanishes at zero temperature, in accord with the
third law of thermodynamics. The relation between $S$ and $C$ in
Eq.~(\ref{SwC}) implies the proportionality between $S$ and $c$ at the
critical point (CP), namely, $S=\pi\,c/(3\,\beta)$.  This is
reminiscent of the relation between geometric entropy for a CFT and
central charge found in \cite{HLW}.

The theorem of increase of the relative entropy (\ref{pos}) holds 
on a finite geometry and guarantees that $S_{\rm
  rel}(\l,\beta,\Lambda)$ increases with $\l$.
We calculate the relative entropy substituting the value of $W=-\ln
Z/(\beta\,L) = F/L$ according to Eq.~(\ref{fse}):
\begin{eqnarray}
S_{\rm rel}(\l,\beta,\Lambda) = W(\l,\beta,\Lambda) - W(0,\beta,\Lambda) - 
\l{\partial W(\l,\beta,\Lambda)\over \partial \l} = \nonumber\\ 
S_{\rm rel}(\l,\Lambda) 
+{1\over \beta^2}\left(C- C(0) - \l{\partial C\over \partial\l}\right)= 
S_{\rm rel}(\l,\Lambda) + {\pi\,c\over 6\,\beta^2} - 
{1\over \beta^2}\left(C - {x\over y}\,{d C\over dx}\right),   
\label{StoS}
\end{eqnarray}
where $S_{\rm rel}(\l,\Lambda) = \lim_{\beta \rightarrow \infty}S_{\rm
  rel}(\l,\beta,\Lambda)$ is the bulk relative entropy. If this entropy is
universal we have shown that it takes the form $S_{\rm rel}(\l) =
B\,\l^{2/y}$.  Then the presence of the scale $\beta$ allows
us to define a dimensionless relative entropy, 
\be 
{\cal C}(x) =
\beta^2\,S_{\rm rel}(\l,\beta) = {\pi\,c\over 6} + B\,x^{2} -\left(C -
  {x\over y}\,{d C\over dx}\right).
\label{cC}
\ee
Furthermore, the monotonicity theorem adopts a dimensionless form,
\be
x{d {\cal C}\over d x} = 
{\beta^2\over y}\int\!\! d^2z\,
\langle :\!\Theta(z)\!:\,:\!\Theta(0)\!:\rangle,   
\ee
Since derivatives with respect to $x$ are equivalent to derivatives
with respect to $\beta$, $\C$ embodies RG irreversibility, in the
manner of Zamolodchikov's theorem \cite{ZamoC}.  Although $\C(0)=0$,
we can redefine it such that it is proportional to the central charge
$c$ at the CP by substracting the constant term $\pi c/6$ from both
sides of Eq.~(\ref{cC}), enhancing the similarity with Zamolodchikov's
$c$ function.  We could say that it also plays the r\^ole of an
off-critical ``central charge".  {}From (\ref{cC}) it is clear that
${\cal C}(x)$ has a bulk part proportional to $x^{2}$ and a
finite-size correction, expressed in terms of $C(x)$.  As $x
\rightarrow \infty$, $C(x)$ tends to a finite limit and so does the
finite-size part of ${\cal C}(x)$. Hence, in the low-temperature limit
$x \rightarrow \infty$ the bulk part dominates, ${\cal C}(x)\approx
B\,x^2$, so that $\C(x)$ diverges, unless $B=0$. If the relative
entropy is not universal, we can nevertheless define a dimensionless
relative entropy but then as a function of two variables, namely, ${\cal
  C}(x,\beta\Lambda)$.  Since we must have that $\beta\,\Lambda \gg
1$, monotonicity still holds for moderate values of $x$.

In parallel with the relative entropy, now it is natural to consider 
the behaviour of the absolute $1d$ quantum 
entropy $S$ with respect to $\beta$:
\begin{equation}
{\partial S\over \partial\beta} = 
{\partial \over \partial\beta}(\beta\,E- \beta\,F) = 
\beta\,{\partial E\over \partial\beta} = 
\beta\,{\partial^2(\beta\,F)\over \partial\beta^2}.
\end{equation}
We have again monotonicity, for $\beta\,F$ is a convex function of
$\beta$, as deduced from the expression of its second derivative as
the average $-\langle (H - \langle H \rangle)^2 \rangle$. Unlike the
monotonicity of the $2d$ relative entropy, Eq.~(\ref{pos}), here $H$
is the {\em total} Hamiltonian, that is, including the critical part
$H^*$---e.g., the kinetic term $H^*=\int(\partial \phi)^2/2$.  This
monotonicity is in principle unrelated to the monotonicity of
$S_{\rm rel}$ with respect to the coupling constant. Thus it 
allows us to define a different monotonic dimensionless function,
\be
{\tilde\C}(x) = {S\over L\,\l^{1/y}}= -{2\,C\over x} + {d C\over dx}.
\ee
At the critical point $S/L=\pi\,c/(3\,\beta)$, implying that
${\tilde\C}(x)$ diverges linearly at $x=0$, whereas ${\cal C}(0) = 0$.
On the other hand, as the temperature is lowered ($x\rightarrow\infty$) 
$\tilde\C(x)$ decays to zero.

We see that there are several quantities that can be related at a RG
fixed point but have a different physical origin and clearly differ
away from it.  The quantity which has been more prominent in the
literature is the finite size correction to the free energy $C(x)$. It
was proposed as a monotonic function in Refs.~\cite{BoyHo,CaNFr}.  It
has sometimes been related to the dimensionless quantity
$3\,\beta^2\,\langle T \rangle/\pi$, which gives the central charge
$c$ at the fixed point.  To clarify this question we prove here that
this expectation value is instead related to the $1d$ quantum entropy
$S$, showing on the way the general relation of expectation values of
stress tensor components with thermodynamic quantities. Let us
consider the expectation values of the complex components of the
stress tensor, $\Theta := T_a^a$ and $T :=
T_{11}-T_{22}-2\,i\,T_{12}$, on the cylinder geometry.  We have the
equalities $$E/L = \langle T_{11}\rangle, \quad F/L = \langle
T_{22}\rangle,$$
which come from the definition of the stress tensor
and are completely general. One deduces that
\be
S/(L \beta) = \langle T_{11} - T_{22}\rangle  = \langle T\rangle,
\ee 
which generalizes the standard relation $F/L = (-1/2)\,\langle
T\rangle$ \cite{BCN,Aff}, actually only valid at the fixed point. However, 
the monotonic function
$\tilde{\cal C}= (\beta/\l^{1/y})\,\langle T\rangle$ is related with the
expectation value $\langle T \rangle$. In fact,
\be
\langle T\rangle = -{2\over\beta^2}\left(C - {x\over 2}{d C\over dx}\right),
\ee
containing the term $x\,dC/dx$, which vanishes at the fixed point. 

The coupling may have been understood in all the above as taking the
statistical system off criticality. However, nothing in the arguments
above requires that for $\l \neq 0$ the correlation length be finite.
Actually, we can well envisage the situation in which a coupling of a
system at a multicritical point is such that the coupled system is
still critical.  This situation is described in field theory as a {\em
  massless flow}, which causes the system to undergo a crossover
ending at another non-trivial fixed point of the RG.  However, we will
only study here massive flows, with a finite correlation length and
hence a mass parameter $m$. In free theories, as considered in
\cite{I}, $m$ is the mass of the particles, bosons or fermions. In
interacting theories there is a mass spectrum, which can be deduced
from the long distance behaviour of the two-point correlation
function. We will be considering theories soluble with the TBA, which
directly renders the mass spectrum. One may then select the lowest
mass of the TBA spectrum and define the dimensionless variable as
$x=\beta\,m$.  In massive theories the function $C(x)$ vanishes
exponentially as $x\rightarrow \infty$, and so do the entropic
functions ${\cal C}(x)$ and $\tilde{\cal C}(x)$.

\section{Finite size thermodynamics for free field models}

\subsection{The continuum limit of the lattice Gaussian and Ising models}

The $2d$ Gaussian model on a square lattice with thermal
coupling constant $\hbox{\ss}$ is exactly soluble,\footnote{The
  thermal coupling constan of a $2d$ lattice model is of course the
  inverse $2d$ {\em temperature}. Since we shall be using throughout
  the corresponding $1d$ temperature, $\beta =1/T$, we avoid mentioning a
  $2d$ temperature and use the notation $\hbox{\ss}$ for the $2d$
  coupling constant.}  yielding
\begin{equation}
W(\hbox{\ss})={1\over2}\int\limits_{-\pi}^{\pi}{d^2k\over{(2\pi)}^2}\,
\ln\left[1 - 2\,\hbox{\ss}\,(\cos k_x+\cos k_y)\right]    \label{WG}
\end{equation}
per site \cite{Parisi}. It has a CP for $\hbox{\ss}_c=1/4$.  The
continuum limit is performed by redefining wave vector as $k=a\,p$,
$a$ being the lattice spacing, and considering $W$ per unit area.
Although $k$ belongs to a Brillouin zone, in the continuum limit $p$
runs over the domain, $-\Lambda<p_{x},p_{y}<\Lambda$ ($\Lambda \sim
\pi/a$), which becomes the entire plane as $\Lambda\rightarrow\infty$.
In the continuum limit we have the field theory of free bosonic
particles of mass $m$ such that
\be
m^2\,a^2 = 16\,(\hbox{\ss}_c-\hbox{\ss}), \label{Gcoup}
\ee
so that $y=2$ and the coupling is $r =m^2$, omitting an irrelevant
proportionality constant.

The relative entropy per unit area of the Gaussian model is best 
calculated with field theory methods, for example, 
using dimensional regularization \cite{I-OC}. 
It can be expressed as 
$$ S_{\rm rel}=
\frac{\Gamma[(4-d)/2]}{(4\pi)^{d/2}\,d}\,r^{d/2},$$ which in $d=2$ yields
\begin{equation}
S_{\rm rel}= \frac{r}{8\pi},
\end{equation}
However, it
is more illustrative to start with the expression of the cutoff
logarithm of the partition function per unit area
\begin{equation}
W[r,\Lambda] \equiv -\ln Z[r,\Lambda] =
{1\over2}\int\limits_{0}^{\Lambda}{d^2p\over
{(2\pi)}^2}\,\ln{p^2+r\over\Lambda^2},
\label{W}
\end{equation}
which can be integrated exactly and yields
\begin{equation}
W[r,\Lambda] = {1\over 2\pi} {\Lambda^2\over 4} \left\{-1-
{r\over \Lambda^2} \ln {r\over \Lambda^2} + 
\left(1+{r\over \Lambda^2}\right) \ln\left(1+{r\over \Lambda^2}\right)\right\}.
\end{equation}
Naturally, it is UV divergent. For large $\Lambda$ it becomes
\begin{equation}
W[r,\Lambda] = {1\over 8\pi} \left\{-{\Lambda^2} + {r} \ln {\Lambda^2\over r} 
+ {r} + {\rm O}(\Lambda^{-2})\right\},
\label{Wexpa}
\end{equation}
exhibiting a quadratic and a logarithmic divergence. 

Recalling the discussion on the general structure of $W$ in the
previous section, we see that we are in the case of logarithmic
corrections to a pure scaling form.  Nevertheless, it is easily derived
that in the present case all the divergences in
$\Lambda$ cancel in the relative entropy, yielding in the infinite
cutoff limit
\begin{equation}
S_{\rm rel} = W(r) - W(0) - r{dW\over dr} = \frac{r}{8\pi},  \label{Srel}
\end{equation}
in accord with the dimensional regularization result. To be precise,
in this cutoff regularization the quadratic divergence cancels by the
substraction of $W(0)$ and the logarithmic divergence by the Legendre
transform, while in dimensional (or analytic) regularization the
quadratic divergence does not appear but there is a pole in $W$,
equivalent to the logarithmic term in $\Lambda$, that cancels in
$S_{\rm rel}$.

Another interesting and exactly soluble example is the $2d$ 
Ising model on a square lattice, with
\begin{equation}
W(\hbox{\ss}) =-{1\over 2}\int\limits_{-\pi}^\pi {d^2k\over{{(2\pi)}^2}}\,
\ln\left[\cosh^2(2\,\hbox{\ss})- 
\sinh(2\,\hbox{\ss})(\cos k_x+\cos k_y) \right] - \ln 2
\label{Wps}
\end{equation}
per lattice site \cite{Onsager}.
The critical point occurs for the value of $\hbox{\ss}$ 
such that the
argument of the logarithm vanishes when $k_x, k_y \rightarrow 0$, namely, when
\begin{equation}
f(\hbox{\ss}) := \cosh^2(2\,\hbox{\ss})-2\,\sinh(2\,\hbox{\ss}) = 0,
\end{equation}
with solution $\hbox{\ss}_c = {{\rm ArcSinh}(1)}/2 = 
\ln({\sqrt{2}}+1)/2 \approx 0.440687$. The expansion of $f(\hbox{\ss})$ near
$\hbox{\ss}_c$ yields
$$
f(\hbox{\ss}) =  8\,{{\left(\hbox{\ss}  -  \hbox{\ss}_c\right) }^2} + 
   8\,{\sqrt{2}}\,{{\left( \hbox{\ss} - \hbox{\ss}_c \right) }^3} + 
   {{{\rm O}(\hbox{\ss} - \hbox{\ss}_c)}^4}.
$$
If we define $m$ by 
\be
m^2\,a^2 = 16\,(\hbox{\ss}-\hbox{\ss}_c)^2   \label{Icoup}
\ee
and redefine the momentum $k$ as $k = a\,p$, with $a$ the lattice spacing, we
obtain near the critical point that
\begin{equation}
W(\hbox{\ss}) = -{a^2\over2}\int\limits_{-\pi/a}^{\pi/a}{d^2p\over
{(2\pi)}^2}\,\ln{\left[(p^2+m^2)\,a^2/2\right]}.
\label{Wpv}
\end{equation}
In other words, the corresponding field theory is described by a $W$
per unit area given by minus that in Eq.~(\ref{W}). It agrees with
the well known description of this model in terms of a free Majorana
fermion theory.  However, the relative entropy is not minus that of
the Gaussian model, since now the coupling constant is proportional to
$m$ instead of being $r=m^2$, since Eq.~(\ref{Icoup}) implies that
$y=1$.  One obtains
\begin{equation}
S_{\rm rel}(r) = W(r) - W(0) - m{dW\over dm} = -\frac{m^2}{8\pi}\,
(1+\ln {m^2\over \Lambda^2}).  \label{SrelI}
\end{equation}
It diverges in the limit of infinite cutoff, which cannot be removed
to obtain a universal value.  Nevertheless, for $m \ll \Lambda$, where
the field theory makes sense, the relative entropy in (\ref{SrelI}) is
monotonic. This is not surprising because it coincides near the CP
with the exact relative entropy of the square-lattice Ising model, 
represented in \cite{I-OC}.

\subsection{Derivation of finite-size quantities}

The expression of $W$ on a lattice of finite size $L_1 \times L_2$ is
obtained by replacing the integrals in (\ref{WG}) or (\ref{Wps}) with
sums over discrete momenta with step ${2\pi/L_1}$ and ${2\pi/L_2}$.
When $L_1, L_2 \gg a$ we approach the thermodynamic limit and the sums
become integrals plus some finite-size corrections.  However, the
double limit $L_1, L_2 \rightarrow \infty$ is complicated to study,
and it is better to consider finite-size effects only in one
direction.  Alternatively, it is sometimes convenient to consider a
non-symmetrical lattice with different coupling constants in the
horizontal and vertical directions. In particular, the quantum $1d$
Gaussian or Ising models on a chain of sites can be obtained as the
extreme anisotropic limit of the $2d$ Gaussian or Ising models
\cite{Suzu,FradSuss}.  The CP is still where the correlation length
diverges but now correlations are calculated only between horizontal
spins.  Now the partition function is $Z = {\rm Tr}\,e^{-\beta\,H}$,
which can be represented in the continuum limit as a functional
integral on $\IR \times S^1$ with $\beta=1/T$ the length of the
compact dimension. It may be good to recall that here $\beta$ has no
relation with the coupling constant, unlike $\hbox{\ss}$ in the
classical $2d$ models above, and plays instead the r\^ole of RG
parameter.

Let us first consider the specific ground-state energy of the $1d$
lattice system.  For the Gaussian model in the continuum limit it is
given by
\begin{equation}
e_0 = \int\limits_{0}^{\Lambda}{dp\over{2\pi}}\,\sqrt{p^2+m^2} =
  {{1}\over 4\pi}
    \left[{\Lambda}\,{\sqrt{{{{\Lambda}}^2} + {m^2}}} - 
      {m^2}\,\log m + 
      {m^2}\,\log ({\Lambda} + {\sqrt{{{{\Lambda}}^2} + {m^2}}})\right].   
 \label{e0}
\end{equation}
When $\Lambda \rightarrow \infty$ the leading terms are
\begin{equation}
e_0 = {1\over 8\pi} \left\{2\,\Lambda^2 + 
2\,{m^2} \ln {2\,\Lambda\over m} 
+ {m^2} + {\rm O}(\Lambda^{-2})\right\}.    \label{e02}
\end{equation}
It is quadratically divergent. The logarithmic divergence is
universal, that is, independent of the regularization method, and
corresponds to the logarithmic divergence of $W[r]$,
Eq.~(\ref{Wexpa}).  However, note that the $\Lambda^2$ and $m^2$ terms
are non universal and their coefficients change from $-1$ and $1$ in
(\ref{Wexpa}) to $2$ and $2 \ln 2 + 1$, respectively.  We shall show
below that the specific ground-state energy of the Ising model is
given by the same formula, except for an overall minus sign,
in agreement with its free energy in Eq.~(\ref{Wpv}).

In order to compute finite-size effects we first consider the
behaviour of the ground state energy on a segment of length $L$ at
zero temperature, connected with the well-known Casimir effect.  It
provides the finite-size correction $C(x)$ that we need. To see this,
let us take the specific $W$, according to Eq.~(\ref{fse}),
\begin{equation}
{-\ln Z\over L\,\beta} =
e_0(\Lambda,m) + {C(\beta\,m)\over\beta^2},  
\end{equation}
and interchange the r\^oles of $L$ and $\beta$: we have that at
low-temperature
\begin{equation}
{-\ln Z\over L\,\beta} =  
e_0(\Lambda,m) + {C(L\,m)\over L^2}.  
\end{equation}
Since $E = -{\partial\ln Z/\partial\beta}$, this formula also gives
the specific ground-state energy on a segment of length $L$ at zero
temperature.  The term proportional to $L$ is the bulk ground-state
energy considered above and the finite-size correction is the Casimir
energy.  In other words, the Casimir energy provides the universal
function $C(m\,\beta)$.

\subsubsection{Direct calculation of the Casimir energy}

The Gaussian model with periodic boundary condition has a ground state energy
\begin{equation}
E_0 ={1\over 2}\, \sum_{n=-\infty}^{\infty}
\sqrt{\left({2\pi\,n\over L}\right)^2+m^2}
= {m\over 2}+ \sum_{n=1}^{\infty}\sqrt{\left({2\pi\,n\over L}\right)^2+m^2}. 
\label{E0}
\end{equation}
When $L \rightarrow \infty$ one recovers the continuum integral of 
Eq.~(\ref{e0}). However, if we are interested in the vicinity of 
the critical theory we may consider the limit $L \rightarrow \infty$ 
but with $m\,L$ small. This limit is known to provide a method 
to calculate the CFT central charge. Then the series can be evaluated 
by expanding the square root in powers of $m\,L$ and interchanging the sums.
We obtain
\begin{eqnarray}
E_0 = {m\over 2}+ {2\pi\over L}\, \sum_{l=0}^{\infty} 
\left(\!\!\begin{array}{c}1/2\\l\end{array}\!\!\right) 
\left({m\,L\over 2\pi}\right)^{2l} \zeta(2l-1) =
\nonumber\\  
{m\over 2}+ {2\pi\over L} \left[ \zeta(-1) + 
{1\over 2}\,\left({m\,L\over 2\pi}\right)^{2} \zeta(1) -
{1\over 8}\,\left({m\,L\over 2\pi}\right)^{4} \zeta(3) + \cdots \right].
\label{EG}
\end{eqnarray}
Since $E_0$ is divergent, the result amounts to a {\em zeta-function}
regularization of it. The first term, with $\zeta(-1) = -1/12$, yields
$c=1$. The next term, proportional to $L$, accounts for the bulk term
$e_0$. Despite the regularization, it is still divergent, since
$\zeta(z)$ has a simple pole at $z=1$. This pole is equivalent to a
logarithmic divergence in regularizations with a UV cutoff, as
generally happens when comparing analytic with cutoff regularizations.
The way to realize it for this case is to restrict the sum $\zeta(1) =
\sum_1^\infty (1/n)$ up to some large number $N$.  Then
\begin{equation}
\zeta(1) \simeq \sum_1^N {1\over n} = \log N + \gamma+{\rm
  O}({1\over N}).
\label{zeta1}
\end{equation}
The connection with the regularization provided by
considering the system on a discrete chain of spacing $a$ can be made
taking $N = L/a$, the number of sites.  An alternative procedure of
regularization is first to segregate the divergent bulk part, with the
form (\ref{e0}), from the finite-size corrections by using the
Euler-MacLaurin formula (appendix A).

The Ising model on a closed chain is amenable to an analogous treatment. 
Its ground 
state energy for $T > T_c$ is like (\ref{E0}) but with negative sign and
with wavenumbers that are odd powers of $\pi/L$ \cite{SML}
\begin{equation}
E_0 =- {1\over 2}\, \sum_{n=-\infty}^{\infty}
\sqrt{\left({(2\,n+1)\,\pi\over L}\right)^2+m^2}= 
-\sum_{n=0}^{\infty}\sqrt{\left({(2\,n+1)\,\pi\over L}\right)^2+m^2}. 
\label{E0I}
\end{equation}
The expansion in powers of $m\,L$ yields
\begin{eqnarray}
E_0 =  {2\pi\over L}\, \sum_{l=0}^{\infty} 
\left(\!\!\begin{array}{c}1/2\\l\end{array}\!\!\right) 
\left({m\,L\over 2\pi}\right)^{2l} \left(1-2^{2\,l-1}\right) \zeta(2l-1) =
\nonumber\\  
{2\pi\over L} \left[{1\over 2}\, \zeta(-1) - 
{1\over 2}\left({m\,L\over 2\pi}\right)^{2} \zeta(1) +
{7\over 8}\left({m\,L\over 2\pi}\right)^{4} \zeta(3) + \cdots \right].
\label{EI}
\end{eqnarray}
The central charge is $c=1/2$ and the bulk term is minus that of 
the Gaussian model.

\subsubsection{Thermodynamic calculation of finite-size effects}

Now we concern ourselves with the deviation of the energy at non-zero
temperature from the ground-state energy or, in other words, the
finite-size $\beta$ correction to the free energy. For the Gaussian
model it can be expressed as the free energy of the ideal Bose gas
constituted by the elementary excitations,
\begin{equation}
\beta\,{F\over L} = e_0\,\beta + \int\limits_{-\infty}^\infty {dp\over 2\pi} 
\ln\left(1- e^{-\beta\,\epsilon(p)}\right),
\label{intF}
\end{equation}
where the one-particle energy is $\epsilon(p) = \sqrt{p^2+m^2}$.  This
formula can also be obtained by an explicit calculation of the
finite-size corrections in the $2d$ lattice model \cite{NaOC}. When
$m=0$ it can be used to calculate the central charge \cite{Aff}.
Nevertheless, an expansion in powers of $m^2$ is not advisable: The
ensuing integral at the next order is IR divergent; that is to say,
the expression (\ref{intF}) is non analytic at $m^2=0$. Fortunately,
the integral can be computed by changing the integration variable to
$\epsilon$ and expanding the logarithm in powers of
$e^{-\beta\,\epsilon}$.  We obtain
\begin{equation}
\beta\,{F\over L} = e_0\,\beta 
-{m\over\pi} \sum_{n=1}^\infty {1\over n}\,K_1(n\,m\,\beta), \label{FK}
\end{equation}
where $K_1(x)$ is a modified Bessel function of the second kind. 

We now define the dimensionless quantity $x= m\,\beta$ and perform  
a small-$x$ expansion, which yields
\begin{eqnarray}
\beta\,{F\over L} = e_0\,\beta -{1\over\pi\,\beta}\,\left\{\zeta(2) - 
{\pi\over 2}\,x -
{x^2\over 4}\left(\ln{x\over 4\pi}+\gamma-{1\over 2}\right) +
{\rm O}(x^4)\right\} \\= e_0\,\beta -{\zeta(2)\over\pi\,\beta} +{m\over 2} +
\beta\,{m^2\over 4\pi}\left(\ln{m\,\beta\over 4\pi}+\gamma-{1\over 2}\right) +
{\rm O}(m^4),
\label{FG}
\end{eqnarray}
where ${\rm O}(x^4)$ denotes an analytic remainder of fourth order.
The term $\zeta(2) = \pi^2/6$ gives the usual $m=0$ part and
central charge $c=1$.  The non-analyticity in $m^2$ of the integral for $F$
(\ref{intF}) manifests itself in the appearance of the $m/2$
and logarithmic terms.  The former also appears as a zero mode contribution
in the Casimir energy (\ref{E0}). The full $x$-power series is obtained
as follows: The series of Bessel functions (\ref{FK}) is slowly
convergent and one may apply to it a Mellin transform to convert it
into a rapidly convergent one \cite{MacFar}.  Fortunately, its Mellin
transform is a power series of $x$.  Furthermore, after replacing
$\beta$ with $L$ it coincides with Eq.~(\ref{EG}) from $l=2$ onwards.
Of course, this should be expected on the grounds of symmetry on a
torus under interchange of its sides $L$ and $\beta$, that is, modular
symmetry. Incidentally, the exact free energy on the torus can be
computed with some more sophisticated mathematics \cite{ItDr,NaOC}.
Its two cylinder limits yield Eq.~(\ref{EG}) or Eq.~(\ref{FG}).
Nevertheless, to show the modular invariance of the exact expression
on the torus is not easy: It can be done performing its expansion in
powers of the dimensionless modular invariant parameter $m^2\,A$, with
$A$ the area of the torus, but it is very laborious.

It is interesting to relate the logarithmic term in the expansion of
$F$ (\ref{FG}) with the Casimir energy calculated in the previous
subsection.  It was remarked there that the $\zeta(1)$ divergence can
be interpreted as a logarithmic divergence in the cutoff. Adding the
logarithmic terms in $e_0$ and $C(x)$ one obtains
\be
{x^2\over 4\pi}\left(\ln{\Lambda\,\beta\over 2\pi}+\gamma\right) \approx 
{x^2\over 4\pi}\,\log N,   \label{bulklog}
\ee
where $N$ is the number of lattice sites in the time direction. We see
that it is equivalent to the $\zeta(1)$ divergence (\ref{zeta1}) under
the interchange $\beta\leftrightarrow L$.

Now we can calculate the specific entropy
\begin{equation}
{S\over L} = {\pi\over 3\,\beta} - {1\over 2}\,m + \beta\,{m^2\over 4\pi} 
+ {\rm O}(m^4).
\label{SG}
\end{equation}
It has no IR singularity at $m \rightarrow 0$ as opposed to the free
energy or the energy. The last term is just twice the relative
entropy of a box of size $L\,\beta$, Eq.~(\ref{Srel}), times $\beta$.
In more generality, for the Gaussian model there is a relation between both types of entropy, namely,  
\begin{eqnarray}
S_{\rm rel}(r,\beta) = W(r,\beta) - W(0,\beta) - 
r{\partial W(r,\beta)\over \partial r} = S_{\rm rel}(r)\nonumber\\
\mbox{}+{1\over \beta^2}\left(C- C(0) - r{\partial C\over \partial r}\right) 
= \frac{r}{8\pi} - {S\over 2L\beta} + {\pi\over 6\,\beta^2}.   
\label{StoSG}
\end{eqnarray}
For other models there is no direct relation between the 
$1d$ entropy and the $2d$ relative entropy. 

The derivation of thermodynamic quantities for the Ising model is analogous. 
The free energy is that of an ideal Fermi gas
\begin{equation}
\beta\,{F\over L} = e_0\,\beta -\int\limits_{-\infty}^\infty {dp\over 2\pi} 
\ln\left(1+ e^{-\beta\,\epsilon(p)}\right) = e_0\, \beta + 
{m\over\pi} \sum_{n=1}^\infty {(-)^n\over n}\,K_1(n\,m\,\beta)
\label{intFI}
\end{equation}
where the one-particle spectrum close to the critical point is again 
$\epsilon(p) = \sqrt{p^2+m^2}$. This integral is computed like the 
bosonic one. The small-$m$ expansion yields
\begin{eqnarray}
\beta\,{F\over L} = e_0\,\beta 
-{1\over\pi\,\beta}\,\left\{{1\over 2}\,\zeta(2) + 
{x^2\over 4}\left(\ln{x\over 4\pi}+\gamma-{1\over 2}\right) +
{\rm O}(x^4)\right\} \nonumber\\= e_0\,\beta-{\zeta(2)\over2\pi\,\beta}  -
\beta\,{m^2\over 4\pi}\left(\ln{m\,\beta\over 4\pi}+\gamma-{1\over 2}\right) +
{\rm O}(m^4).
\label{FI}
\end{eqnarray}
As well as for the Gaussian model, it is possible to obtain the
$x$-power series by the Mellin transform of the series of Bessel
functions (\ref{intFI}).  Similarly, after replacing $\beta$ with $L$
it coincides with Eq.~(\ref{EI}) from $l=2$ onwards.

In this case the specific entropy is
\begin{equation}
{S\over L} = {\pi\over 6\,\beta} - \beta\,{m^2\over 4\pi} 
+ {\rm O}(m^4).
\label{SI}
\end{equation}
For the Ising model the relative entropy is related to the energy,
instead of $S$:
\begin{eqnarray}
S_{\rm rel}(m,\beta) = W(m,\beta) - W(0,\beta) - 
m{\partial W(m,\beta)\over \partial m} =\nonumber\\ S_{\rm rel}(m)
\mbox{}+{1\over \beta^2}\left(C- C(0) - m{\partial C\over \partial m}\right) 
=\nonumber\\  -\frac{m^2}{8\pi}\,(1+\ln {m^2\over \Lambda^2})
- ({E\over L}-e_0) + {\pi\over 12\,\beta^2}. 
\label{StoSI}
\end{eqnarray}

We see that for free theories we can derive explicit formulas for the
free energy---and hence for the entropy,--- as well as perturbative
expansions.  Moreover, both the $1d$ entropy and the $2d$ relative
entropy give rise to monotonic central charges, as we proceed to
study, introducing before for convenience the stress tensor.

\subsection{Expectation values of the stress tensor and entropic C theorems}

The previous section has shown the calculation of finite-size
corrections for various quantities of free models with concepts
pertaining to the $1d$ quantum theories, namely, the lattice Casimir
energy or the statistics of quantum gases. The same results can be
attained with the use of $2d$ Green function techniques, through the
calculation of expectation values of the complex components of the
stress tensor, taking into account their relation with thermodinamic
quantities already remarked. We shall rewrite the monotonic functions
in a suitable way to confirm these relations, hence explaining the
structure of those functions.  We thus start with the expressions of
the expectation values, $\Theta := T_a^a$ and $T :=
T_{11}-T_{22}-2\,i\,T_{12}$, in the cylinder geometry, as derived by
$2d$ Green function techniques (appendix B):
\begin{eqnarray}
\langle \Theta \rangle = 
\pm{m^2\over 2\pi}\left(K_0(0)+ 2\sum_{n=1}^\infty(\pm)^n K_0(n\,m\,\beta)\right),\\
\langle T \rangle = 
\pm{m^2\over 2\pi}\left(K_2(0)+ 2\sum_{n=1}^\infty(\pm)^n K_2(n\,m\,\beta)\right), 
\label{T0}
\end{eqnarray}
with the same sign convention as before.  The modified Bessel
functions are divergent at zero, namely, $K_0(0)$ is logarithmic
divergent and $K_2(0)$ is quadratically divergent. These are UV
divergences, like those already considered for $W$, which can be
removed by normal order.

Using the recursion relations satisfied by the Bessel functions we can 
write the free energy (\ref{FK}) or (\ref{intFI}) as
\begin{eqnarray}
\beta\,{F\over L} &=& e_0\, \beta 
\mp{m^2\,\beta\over 2\pi} \sum_{n=1}^\infty(\pm)^n \left[K_2(n\,m\,\beta)-
K_0(n\,m\,\beta)\right]\nonumber\\ 
&=& -{\beta\over 2} \,\langle T-\Theta 
\rangle =\beta\,\langle T_{22} \rangle,
\label{FT}
\end{eqnarray}
showing its relation with the expectation values of the components of
the stress tensor, an example of the relations obtained at the end of
section 1.  Notice that it implies a definite form for $e_0$, to be
compared with (\ref{Wexpa}) or (\ref{e02}). (See appendix B.)

Similarly, we can calculate
\begin{eqnarray}
{\partial W\over \partial r} = {\partial e_0\over\partial r} \pm 
{1\over 2\pi} \sum_{n=1}^\infty(\pm)^n K_0(n\,m\,\beta) =
{1\over 2\,r}\,\langle\Theta \rangle, \\
{E\over L}  = e_0\pm{m^2\over 2\pi} 
\sum_{n=1}^\infty(\pm)^n \left[K_2(n\,m\,\beta)+ K_0(n\,m\,\beta)\right]\nonumber\\ 
= {1\over 2}\,\langle T +\Theta \rangle =
\langle T_{11} \rangle. 
\end{eqnarray}
The first equation is just a particular case of the expression of the 
derivative of $W$ with respect to $r$ as the expectation value of the
``crossover part" of the action \cite{I-OC}, since $\Theta$ is
proportional to it. Having the values of $W$ and its derivative available 
we further obtain for the Gaussian model that 
\begin{eqnarray}
S_{\rm rel}(r,\beta) =  S_{\rm rel}(r) +{\pi\over 6\,\beta^2}- {r\over 2\pi} 
\sum_{n=1}^\infty K_2(n\,m\,\beta) \nonumber\\ = S_{\rm rel}(r) +
{\pi\over 6\,\beta^2}-{1\over 2} \,\langle :\!T\!: \rangle
\end{eqnarray}
and for the Ising model that 
\begin{eqnarray}
S_{\rm rel}(r,\beta) =  S_{\rm rel}(r) + {\pi\over 12\,\beta^2}+ {r\over 2\pi} 
\sum_{n=1}^\infty(-)^n\, [K_2(n\,m\,\beta)+K_0(n\,m\,\beta)] \nonumber\\ 
= S_{\rm rel}(r) +
{\pi\over 12\,\beta^2}-{1\over 2} \,\langle :\!T+\Theta\!: \rangle.
\end{eqnarray}
We have substituted $S_{\rm rel}(r)$ for $e_0(r) - e_0(0) -
r\,\partial e_0(r)/\partial r$.  One obtains the finite expectation
values of normal-ordered stress tensor components owing to the
subtraction of $W(0,\beta)$.\footnote{One must be careful when
  evaluating ${\partial e_0(r)\over\partial r}$. Since $e_0(r) = \mp
  {m^2\over 2\pi} \left[K_2(0)-K_0(0)\right]$ (appendix B), it may
  seem that $e_0(r) - r\,{\partial e_0(r)\over\partial r} \equiv 0$.
  However, $K_2(0)$ and $K_0(0)$ contain an $m$ dependence, because of
  regularization.}  Using the connection between the $1d$ entropy and
the stress tensor, $S/(L \beta) = \langle :\!T\!:\rangle $, pointed
out at the end of section 2, one can directly obtain $S$.

Thus the dimensionless relative entropies for the Gaussian or
Ising models, respectively, are 
\bea
\C(x)={\pi\over 6} +{x^2\over8\,\pi}
- {x^2\over2\,\pi} \sum_{n=1}^\infty K_2(n\,x),\\
\C(x)={\pi\over 12} -
{x^2\over8\,\pi}\left[1+\ln {x^2\over (\Lambda\,\beta)^2}\right]
+ {x^2\over2\,\pi} \sum_{n=1}^\infty (-)^n\,[K_2(n\,x)+K_0(n\,x)].
\label{CcIsing}
\eea
The other monotonic quantity, $\tilde{\cal C}= S/(L\,m)$, 
is common and can be written as
\be
\tilde{\cal C}(x) =
{x\over\pi} \sum_{n=1}^\infty (\pm)^{n+1}\,K_2(n\,x),
\ee
which is, of course, $\tilde{\cal C}= (\beta/m)\,\langle\, :\!T\!: \,\rangle$, 
according to the expression of $\langle T\rangle$ (\ref{T0}).
Series expansions of $C$, $\C$ and $\tilde\C$ are derived from 
Eq.~(\ref{FG}) or (\ref{FI}). 
Both $\C$ and $\tilde\C$ are plotted in Fig.~1. The Ising model $\C$ is 
for the value $(\Lambda\,\beta)^2 = 10000$. It is useful to recall that 
in general $(\Lambda\,\beta)^2 \sim N$, the number of $2d$ lattice sites 
in a box of side $\beta$. 

\vspace{0.8cm}
\includegraphics[width=7cm]{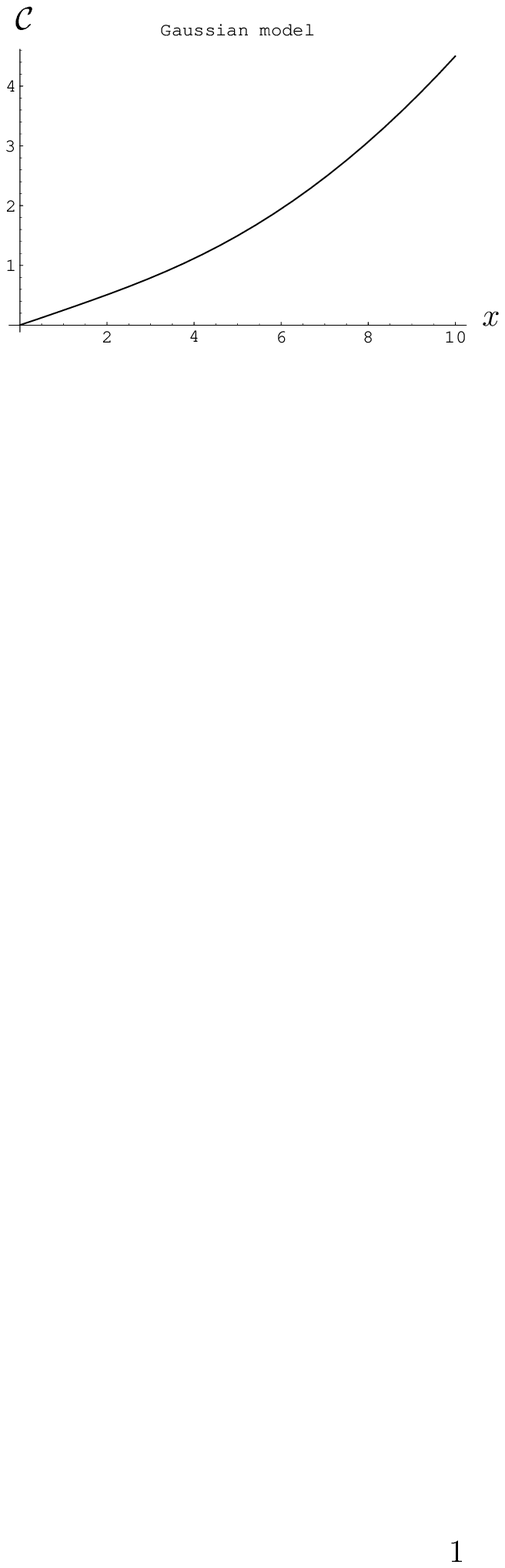}
\includegraphics[width=7cm]{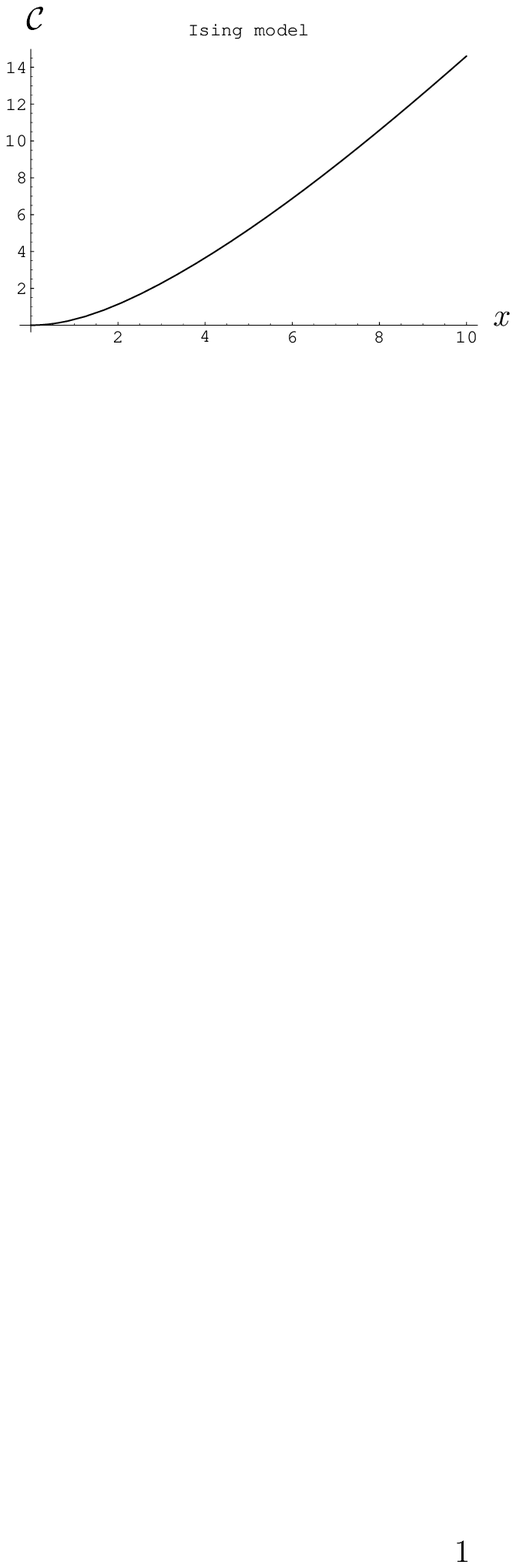}

\includegraphics[width=7cm]{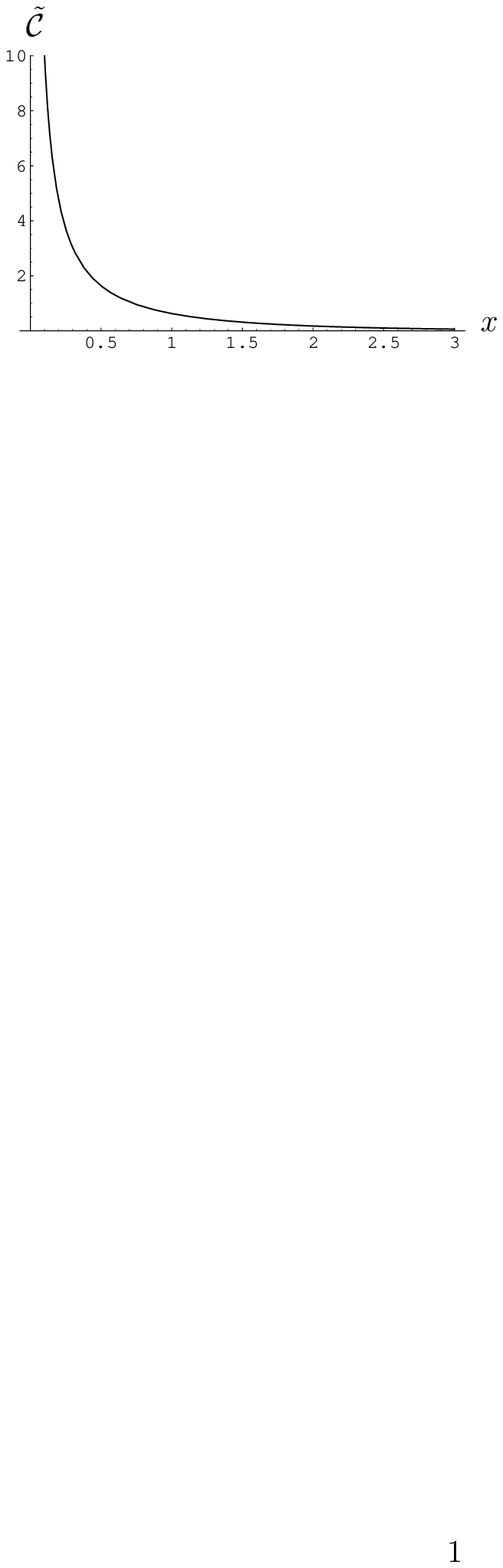}
\includegraphics[width=7cm]{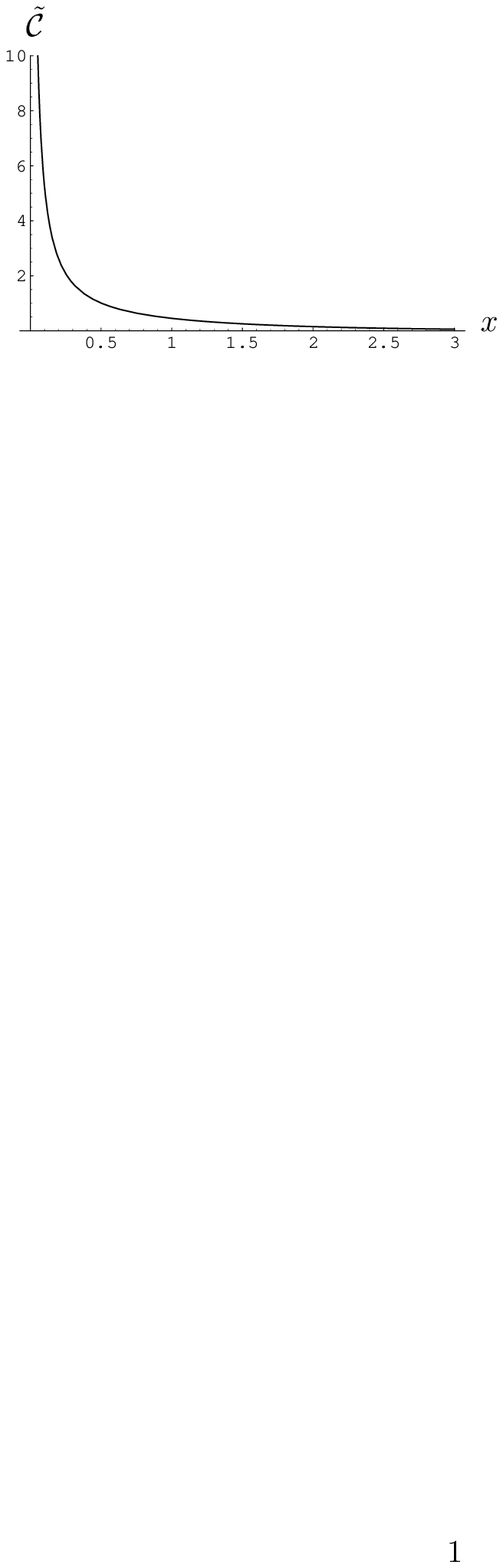}
\vspace{0.5cm}

\centerline{Fig.~1: $\C$ and $\tilde\C$ for 
the Gaussian and Ising models. In the latter model $\C$ is for 
$\Lambda\,\beta = 100$.}

\section{Interacting models}

For interacting models, the free energy is in principle not available
in closed form.  Nevertheless, one can perform its perturbative
expansion.  In two dimensions one can take advantage of the
information provided by the methods of conformal field theory (CFT),
namely, the non-perturbative dimensions and correlations of fields at
the critical point; then one speaks of {\em deformed CFT's}.  For
example, one can perform a perturbative expansion around the critical
point. This approach, called {\em conformal perturbation theory}, is
well suited for the calculation of an entropy relative to the critical
point; the perturbation parameter is $\lambda\,\beta^{y}$.  The
expansions of Eq.~(\ref{FG}) and (\ref{FI}) are instances of it and
can be obtained from the respective $c=1$ or $c=1/2$ CFT \cite{ItDr}.
The logarithmic term in them is due to UV divergencies.  The analysis
of UV divergences is done by examining the behaviour of integrals of
correlators for coincident points. If $1 < y < 2$ there are no UV
divergences in the perturbative expansion and, furthermore, this
expansion is arguably convergent \cite{KlasMel,ConFlu}. In contrast,
when $y<1$ a finite number of terms will diverge. The condition $1 < y
< 2$ agrees with the non-perturbative regularity condition for the
relative entropy found before.  Conformal perturbation theory is
considerably powerful but, at any rate, the perturbative expansion
only converges for a limited range of $\lambda\,\beta^{y}$, while we
are interested in the behaviour of thermodynamical quantities over the
entire range of the coupling constant.

Some 2d models are partially soluble with the thermodynamic Bethe
ansatz (TBA) \cite{YY}. In particular, models for which the
interaction is of purely statistical nature lend themselves to a
derivation of closed expressions for the free energy and entropy
similar to the ones for free field models, albeit more complicated.
Hence, the entropic $C$ theorems can be explicitly verified for them.
The Bethe ansatz assumes a factorized form for the wave functions and
hence an expression for the energy as a sum of contributions of
independent quasi-particle levels, though these quasi-particles have
non-trivial exchange properties.  To determine the structure of these
levels is a complicated business but it dramatically simplifies in the
thermodynamic limit, constituting the basis of the TBA.  This method
is non perturbative in nature and provides thermodynamical quantities
over the entire range of the coupling constant. In principle, the TBA
yields, further to the universal finite-size correction function
$C(x)$, a contribution proportional to $x^{2}$, which is interpreted
as a UV-finite bulk term and therefore has been called {\em universal
  bulk term} \cite{Zamo,KlasMel}. Therefore, in this section we
redefine $C(x)$ to include this universal bulk term.

In spite of the virtues of the TBA approach, the TBA equations
themselves are by no means easy to solve and it is customary to resort
to numerical calculation to obtain the coefficients of the series
expansion in $\beta\,m$. In this sense the TBA approach is not
superior to perturbation theory in its region of convergence, with
which one obtains analytic expressions for these coefficients. It is
only in the case of purely statistical interaction where the TBA
approach is definitely superior, for one can then solve the TBA
equations algebraically. Thus we treat this case first. It applies to
models of Calogero-Sutherland type, which represent the dynamics of
spinons or other non-interacting particles with fractional statistics.
We can calculate the entropy and check that is monotonic with respect
to $\beta$. However, it is beyond our means to calculate expectation
values of the stress tensor, since the expression of the stress tensor
is not available.  In second place, we shall treat the general
interacting integrable case by the numerical solution of the TBA
equations and compare with the results of conformal perturbation
theory.

Last, there is an alternative to conformal perturbation theory or the
TBA that can be applied to any model, namely, numerical finite-size
scaling on a chain \cite{ChriHen}. Like the TBA, by its own nature it
is not limited to a restricted range of $\beta\,m$. However, the
numerical calculations required to obtain similar accuracy to that of
the TBA are prohibitive in practise. Thus this brute-force method is
not actually effective to compute {\em off-critical} quantities and no
use will be made of it here.

\subsection{Models with purely statistical interaction}

The r\^ole of fractional statistics in Condensed Matter Physics, as a
generalization of the regular bosonic or fermionic symmetry properties
under particle exchange, has been long recognized \cite{FraKa}.  Its
modern version has given rise to the concept of anyons.  As it
happens, this type of statistics leads to highly non-trivial
correlations between particles which are difficult to disentangle and
indeed constitute what has been called statistical interaction. The
form of this interaction can be best realized by transforming the
particles to standard fermions or bosons with a peculiar interaction.
Models with purely statistical interaction are usually referred to as
generalized ideal gases \cite{Wu,IsaVi}.  It is customary to consider
the free particles as fermions and parametrize the statistics by a
number $g$, such that the maximum number of particles that can fit in
a single fermion momemtum level is $1/g$.  For no statistical
interaction $g=1$. If $g=1/n$ the single-fermion levels can accomodate
$n$ particles and in the limit $n \rightarrow \infty$ the statistics
becomes bosonic.  Some models with apparently complex interactions can
be transformed into generalized ideal gases, as occurs for the
Calogero-Sutherland models \cite{CaSu} or their lattice version
\cite{Haldane}.

Therefore, models with purely statistical interaction 
are interesting systems and, in addition, 
sufficiently complex to be a suitable benchmark for our irreversibility 
theorem. We start with the spin $SU(2)$ level one 
Wess-Zumino-Witten Lagrangian in the bosonic representation, 
\be
H = \int \!dx\,[(\partial_t\phi)^2 + (\partial_x\phi)^2].
\ee

The spinon field is defined in terms of the bosonic field as $\psi^\pm
= \exp(\pm{i\over \sqrt{2}}\,\phi)$, where the sign stands for spinon
polarization \cite{Haldane,BoLuSa,BerPasSer,Tsvelik}.  It is a free
theory, which can also be expressed as a free fermionic theory, but we
are going to do some non-trivial manipulations on it. First, we
``simplify'' the model by keeping just one spinon polarization, ``+''
say.  The physical way to achieve this is to introduce a very strong
magnetic field. Now we have a {\em semionic} CFT with central charge
$c=3/5$ \cite{SchouPRL,TBA}, which is certainly an interacting theory.
However, its partition function is known, as the total partition
function of the $SU(2)$ level one Wess-Zumino-Witten mode restricted
to vanishing fugacity of ``$-$'' spinons, $z_{-} = 0$,
\cite{BerPasSer}.  The thermodynamic quantities can be obtained with
the help of the TBA.\footnote{Derivation of thermodynamic quantities
  from CFT usually demands a thermodynamic approach in the sense of
  \cite{TBA}, be the TBA or Schoutens' recursion method
  \cite{SchouPRL}.}

The second change consists of the addition of some tunable coupling,
which perturbs the model away from criticality and allows one to probe
the behaviour of the entropy. If we impose that the interaction
remains purely statistical, the only possibility is to give mass to
the semions: We can replace the dispersion relation $\epsilon(p) =
|p|$ with $\epsilon(p) = \sqrt{p^2 + m^2}$. We assume that this
perturbation fulfills the conditions for the application of the
monotonicity theorem (\ref{pos}) and we shall see that it is easily
implemented within the TBA approach and yields expressions which can
be treated by algebraic methods.  As a side remark, note that the
massive relativistic dispersion relation differs from the
non-relativistic one assumed in the original Calogero-Sutherland
model. Therefore, if we want to keep to the physics represented by
this model, we must interpret the mass as a parameter unrelated to the
real semion mass.  For comparison, it can be shown that the $2d$ Dirac
Lagrangian with a mass term appears as a low energy effective
Lagrangian for (non-relativistic) conducting electrons in
one-dimensional metals, but the Dirac mass is actually related to the
electric potential \cite[chapter 13]{Tsvelik}.  Regardless of the
precise physical interpretation, we will consider the theory of
relativistic massive semions as our first interacting field theory to
investigate the properties of the entropy.

\subsubsection{Application of the TBA to the semion gas}

The full power of the TBA shows in the calculation of finite size corrections 
to thermodynamic quantities. One obtains for the critical semion gas 
\cite{TBA,ByFri} 
\be
\,c = 
{\frac{6\,\beta}{\pi^2}}\int\limits_{0}^{\infty}dk\,
{\log \left[{\frac{2 + {\z^2} + \z\,{\sqrt{4 + {\z^2}}}}{2}}\right]}
= {\frac{6}{\pi^2}}\,L\left(\frac{{\sqrt{5}}-1}{2}\right)
= \frac{3}{5},
\ee
with $\zeta = e^{-\beta\,\epsilon(k)}$, $\epsilon(k) = |k|$. 
We give the semions a mass, replacing the dispersion relation 
$\epsilon(k) = |k|$ with $\epsilon(k) = \sqrt{k^2 + m^2}$. Then
\bea
C(\beta\,m) = 
-{\frac{\beta}{\pi}}\int\limits_{0}^{\infty}dk\,
{\log \left[{\frac{2 + {\z^2} + \z\,{\sqrt{4 + {\z^2}}}}{2}}\right]}
= -\frac{2\,\beta}{\pi}\int\limits_{0}^{\infty }dk\, {\rm ArcSinh}%
\left[ \frac{\z}{2}\right] = \nonumber \\
-\frac{2\,\beta}{\pi}\int\limits_{0}^{\infty }dk 
\sum_{n=0}^\infty (-)^n {\frac{\left( 2\,n \right) !}
    {{2^{2\,n}}\,\left( 1 + 2\,n \right) \,{{n!}^2}}}\,
        \left[\frac{\z}{2}\right]^{(2\,n+1)}
= \nonumber \\ -\frac{2\,\beta}{\pi}\,m
\sum_{n=0}^\infty (-)^n {\frac{\left( 2\,n \right) !}
    {{2^{4n+1}}\,\left( 1 + 2\,n \right) \,{{n!}^2}}}\,
        K_1[(2\,n+1)\,m\,\beta].   \label{Csem}
\eea
One can easily obtain its perturbative expansion in powers of $m\,\beta$ by  
expanding first the modified Bessel function,
\bea
K_1[z] &=& {1 \over z}+ \ln\frac{z}{2}\,\frac{z}{2}\sum_{k=0}^\infty
{\left(\frac{z^2}{4}\right)^k \over k!\,(k+1)!} - 
\frac{z}{4}\,\sum_{k=0}^\infty[\psi(k+1)+\psi(k+2)]\,
{\left(\frac{z^2}{4}\right)^k \over k!\,(k+1)!} \\
&=& {\frac{1}{z}} + \frac{z}{2}\,
\left(\ln\frac{z}{2}  +\gamma - \frac{1}{2}\right) + 
  {{\rm O}(z)}^2 ,
\eea
where $\psi(x)$ is the digamma function. 
However, we will content ourselves with extracting 
the bulk part,
\bea
C(x)|_{\rm bulk} &=& -\frac{2\,x}{\pi}
\sum_{n=0}^\infty (-)^n {\frac{\left( 2\,n \right) !}
    {{2^{4n+1}}\,\left( 1 + 2\,n \right) \,{{n!}^2}}}\,
\left[\frac{z}{2}\,
\left(\ln\frac{z}{2}  +\gamma - \frac{1}{2}\right)\right]_{z=(2\,n+1)\,x} 
\nonumber\\
&=& -{{x}^2\over \pi}\left[S_1\,\ln{\frac{x}{2}} + S_2 
+ S_1\,(\gamma -\frac{1}{2})\right] 
= -{{x}^2\over \sqrt{5}\,\pi}\ln{x} + 0.104744\,{x}^2,
\label{CSb}
\eea
where
$$
S_1 = \sum_{n=0}^\infty (-)^n {\frac{\left( 2\,n \right) !}
    {{2^{4n+1}}\,{{n!}^2}}} = {1\over \sqrt{5}},
$$
$$
S_2 = \sum_{n=0}^\infty (-)^n {\frac{\left( 2\,n \right) !}
    {{2^{4n+1}}\,{{n!}^2}}} \ln(2n+1) = -0.0536114.
$$

Corresponding to $g=1/2$ statistics, 
in a segment of length $L$ the semion momenta of 
completely filled single-particle levels 
are $k_n = (2\pi/L)\, |n/2+1/4|$ with $n \in \IZ$.
Hence, the {\em approximate} ground 
state energy is like (\ref{E0I}) but 
with wavenumbers that are odd powers of $\pi/(2\,L)$ \cite{SchouPRL,TBA}, 
\begin{equation}
E_0 =- {1\over 2}\, \sum_{n=-\infty}^{\infty}
\sqrt{\left({(2\,n+1)\,\pi\over 2\,L}\right)^2+m^2}= 
-\sum_{n=0}^{\infty}\sqrt{\left({(2\,n+1)\,\pi\over 2\,L}\right)^2+m^2}. 
\label{E0S}
\end{equation}
Now, an expansion in powers of $m\,L$ would yield a wrong Casimir energy, 
owing to the approximated nature of the ground state energy.
We can however see that the bulk term
\be
{1\over L}\int\limits_{0}^{\infty}{dn}\,\epsilon(n) = 
-2\,\int\limits_{0}^{\infty}{dp\over{2\pi}}\,\sqrt{p^2+m^2}
\label{E0Sb}
\ee
is twice that of fermions, as corresponds to the double average
occupation number of semions.  We can compare the approximate bulk
non-analytic term in (\ref{E0Sb}) with the exact result of the TBA
(\ref{CSb}).  According to the combined form of IR and UV logarithmic
terms (\ref{bulklog}) that we expect from (\ref{E0Sb}),
\be
C(x)|_{\rm bulk} = -{{x}^2\over 2\pi} \ln{x}+ 0.390536\,{x}^2.
\ee
We see that the first term of the sum $S_1$ reproduces the coefficient 
in this approximation, $1/2$, but the total coefficient, 
$1/\sqrt{5} \simeq 0.447214$, is slightly smaller.

The total non-analytic part is now an infinite series, obtained from 
\bea
\frac{2\,x}{\pi}
\sum_{n=0}^\infty (-)^n \,{\frac{\left( 2\,n \right) !}
    {{2^{4n+1}}\,\left( 1 + 2\,n \right) \,{{n!}^2}}}
\left[\ln\frac{z}{2}\,\frac{z}{2}\sum_{k=0}^\infty
{\left(\frac{z^2}{4}\right)^k \over k!\,(k+1)!}\right]_{z=(2\,n+1)\,x}
\nonumber \\= \frac{1}{\pi} \ln\frac{x}{2}\sum_{k=0}^\infty
\sum_{n=0}^\infty (-)^n\, {\frac{\left( 2\,n \right) !\,(2\,n+1)^{2\,k}}
{{2^{4n+1}}\,{{n!}^2}}}\,
{x^{2\,k+2} \over 4^k \,k!\,(k+1)!} + {\rm analytic}\nonumber \\
=\frac{1}{\pi{\sqrt{5}}}\,\ln\frac{x}{2}\,\left({x^2} + \frac{x^4}{25} - 
        \frac{7\,x^6}{750} + \frac{353\,x^8}{225000} - 
        \frac{2651\,x^{10}}{22500000} - \frac{619619\,x^{12}}{16875000000}
\right.  
\nonumber \\ + \left.{{{\rm O}({x^2})}^7}\right) + {\rm analytic}. 
\label{nanal}
\eea
Interestingly, the series coefficients seem to be rational numbers.

The entropic $C$ functions are easily derived from the expression of $C$ 
(\ref{Csem}) 
\be
\C(x) =  {\pi\over 10} - {\frac{ {x^2}}{2\,{\sqrt{5}}\,\pi }}\,  
         \log ({\frac{{x^2}}{(\Lambda\,\beta)^2}})- 
\frac{x^2}{\pi}\sum_{n=0}^\infty (-)^n {\frac{\left( 2\,n \right) !}
    {{2^{4n+1}}\,{{n!}^2}}}\,
        \left\{K_2[(2\,n+1)\,x] +K_0[(2\,n+1)\,x]\right\} .
\ee
and
\be
\tilde\C(x) =\frac{2\,x}{\pi}
\sum_{n=0}^\infty (-)^n {\frac{\left( 2\,n \right) !}
    {{2^{4n+1}}\,{{n!}^2}}}\,K_2[(2\,n+1)\,x].
\ee
They are plotted in Fig.~2. We use again the value 
$\Lambda\,\beta = 100$. 

\vspace{.8cm}
\includegraphics[width=7cm]{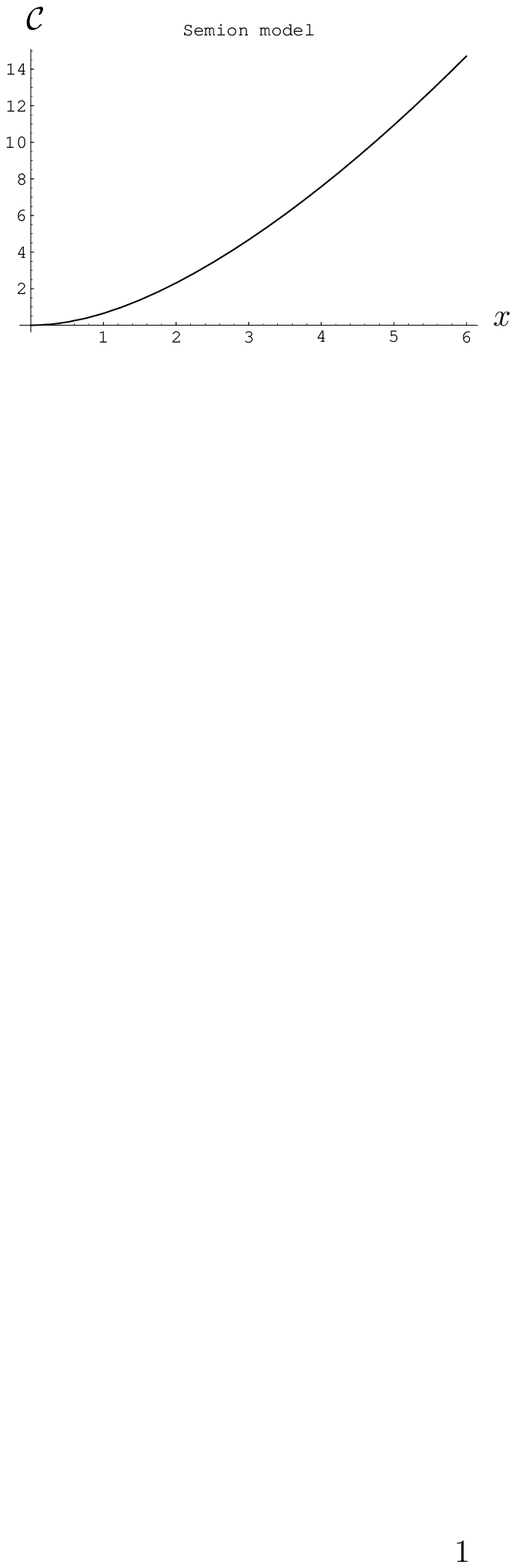}
\includegraphics[width=7cm]{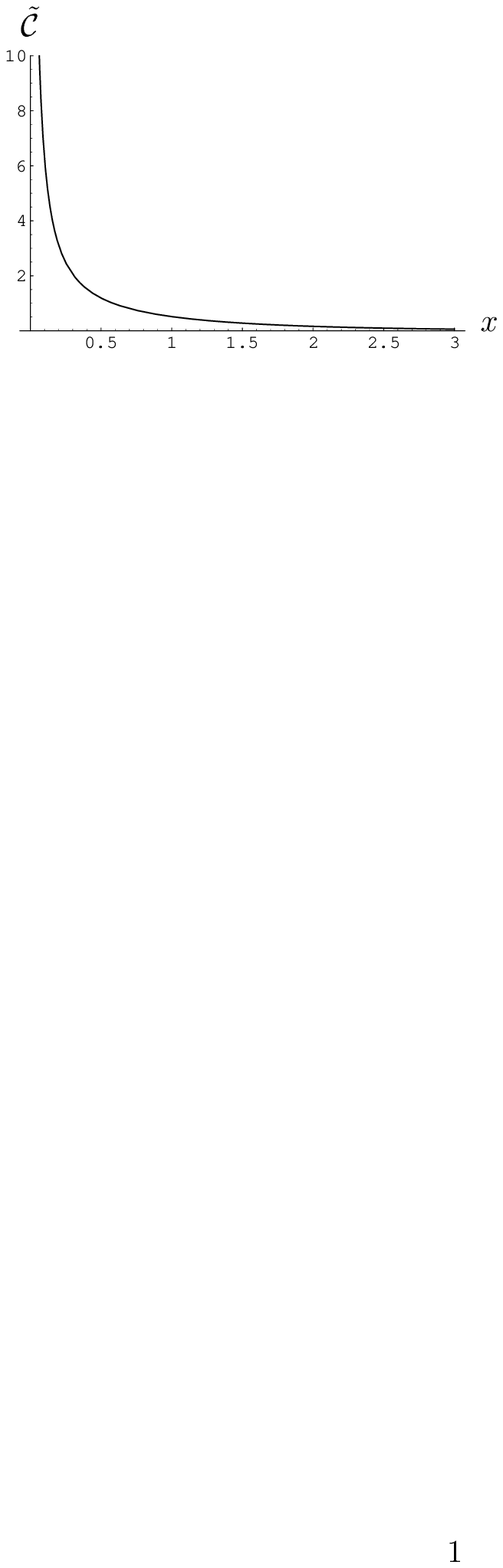}

\centerline{Fig.~2: $\C$ ($\Lambda\,\beta = 100$) and 
$\tilde\C$ for semions.}

\subsection{Deformed two-dimensional conformal field theories}

A general class of theories amenable to derivation of the finite-size
quantities of interest is that of deformed 2d CFT. Since CFT provides
the {\em exact} dimensions of relevant fields, the results of
conformal perturbation theory are more accurate in 2d than those of
ordinary perturbation theory, which on the other hand is plagued with
IR problems.  In addition, many models admit integrable deformations,
in the sense that the existence of an infinite number of conservation
laws forces the $S$-matrix to factorize.  Then the TBA provides a way
to derive thermodynamic quantities.  Many deformed CFT are known to be
integrable and similar methods are applicable to all \cite{Muss},
although their complexity can be considerable for the most
sophisticated models. Therefore, we shall choose one of the simplest
cases. It should be intuitively clear how to generalize the
computation of the entropic quantities to other integrable models.

The natural (and oldest) generalization of the 
Ising model consists of taking a site variable which can take three values 
instead of two, constituting the three-state Potts model. 
It is critical for \ss$_c = \ln(\sqrt{3}+1)/3$ and its 
thermal critical exponent is $\nu = 5/6$, implying that $y=6/5$. 
This model has 
been long known to be integrable and it has been long (but not as long) known 
to be describable in terms of particles with fractional statistics, which 
are generalizations of the Ising fermions and are called parafermions. 
This quasi-fermionic representation is in terms of two conjugate parafermions 
carrying $\IZ_3$ charges +1 or $-1$ and spin 2/3, which are massless 
at the CP point but acquire a mass for $T>T_c$. 
Their interaction is purely statistical at the CP but it is more
complicated off criticality. However, it is still 
integrable and its thermodynamic properties can be found with the TBA. 
It yields
\be
C(\beta\,m) = -\beta\,m \int\limits_{-\infty }^{\infty } \frac{d\theta}{2\,\pi}
\cosh (\theta)\,\ln\left( 1+e^{-\epsilon(\theta)}\right),
\ee
which is apparently similar to the formula for free fermions but now 
$\epsilon(\theta)$ are unknown functions to be determined with 
the TBA equations. The concrete $S$-matrix elements of this model lead to 
the TBA equation
\be
\epsilon(\theta) =\beta\,m\cosh \theta +
\frac{2\sqrt{3}}{\pi }%
\int\limits_{-\infty }^{\infty }d\theta ^{\prime }\frac{\cosh (\theta
-\theta ^{\prime })}{1+2 \cosh 2(\theta -\theta ^{\prime })}\ln 
\left( 1+e^{-\epsilon(\theta-\theta ^{\prime })}\right).   \label{a2}
\ee
It can be solved numerically by an iterative algorithm, yielding a 
set of numbers which can be displayed in a table \cite{Zamo}. 
Hence, according to the general formula (\ref{cC}), 
\be
\C(x) = {\pi\,c\over 6}+
C(x) - {5\,x\over 6}\,{d C(x)\over d x},
\ee
where the bulk relative entropy $S_{\rm rel}(m)$ does not  
appear since it is now implicitly included.    
It derives from 
the universal bulk free energy, which can be calculated exactly, yielding 
$-\sqrt{3}\,x^2/6$ \cite{Zamo}; hence, $S_{\rm rel}(m) = \sqrt{3}\,m^2/9$.
We can calculate the monotonic functions $\C$ $\tilde\C$ numerically as well.
They are plotted in Fig.~3. 

\vspace{.8cm}
\includegraphics[width=7cm]{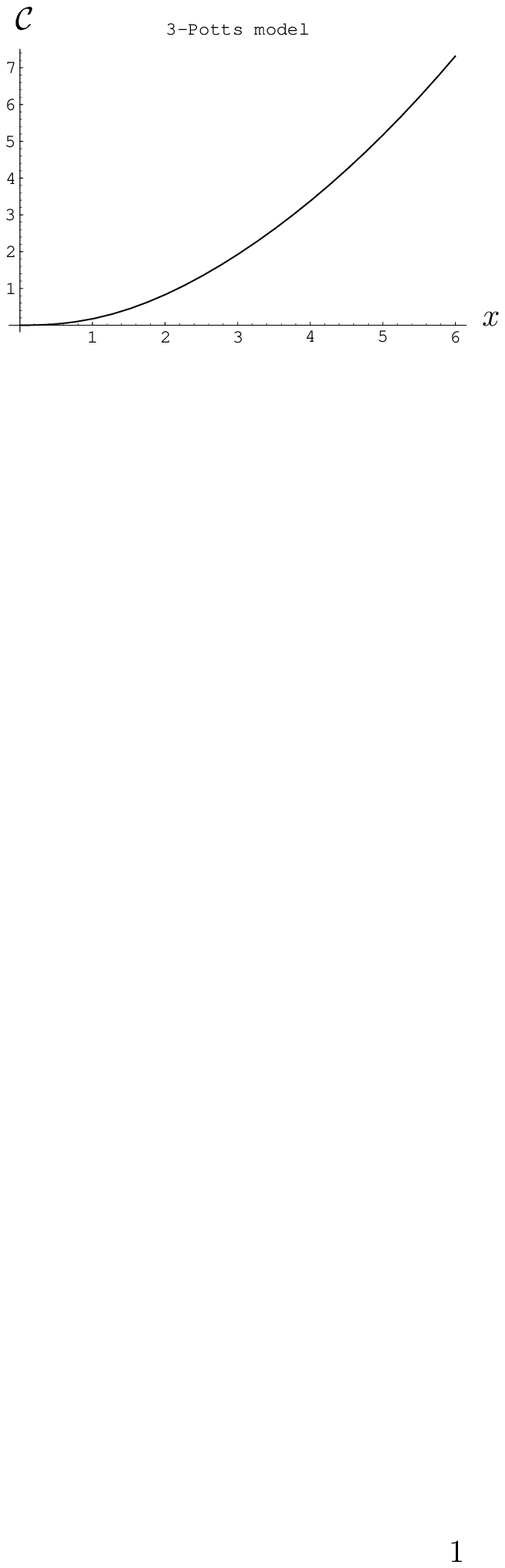}
\includegraphics[width=7cm]{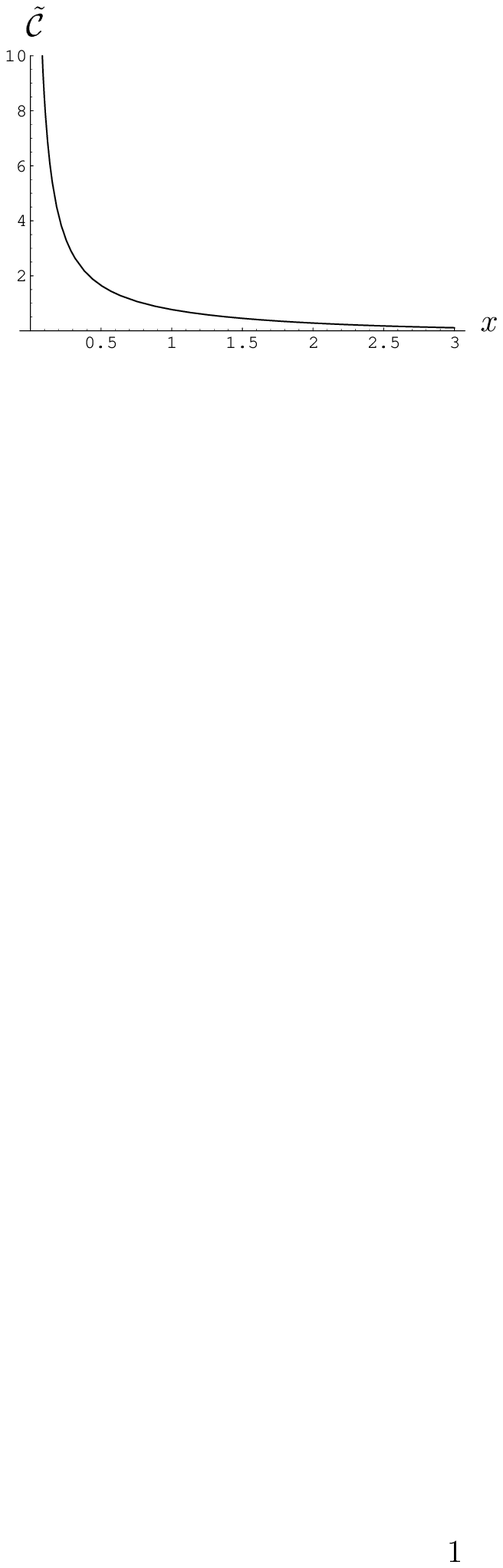}
\vspace{0.5cm}

\centerline{Fig.~3: ${\cal C}$ and $\tilde\C$ for the 3-state Potts model.}
\vspace{1cm}

The previous solution for $C(x)$ can be expressed as 
a power series in $x^{12/5}$ plus a bulk term \cite{Zamo},
\bea
C(x)=-\frac{\pi}{6}\left[
{\frac{4}{5}} - {\frac{{\sqrt{3}}\,{x^2}}{\pi }} + 
   0.339688\,{x^{{\frac{12}{5}}}} - 0.00326095\,{x^{{\frac{24}{5}}}} + 
\right.
\nonumber\\
   0.000114199\,{x^{{\frac{36}{5}}}} - 
   5.11209\,{{10}^{-6}}\,{x^{{\frac{48}{5}}}} + 
   4.01138\,{{10}^{-7}}\,{x^{12}} - 
   1.38691\,{{10}^{-8}}\,{x^{{\frac{72}{5}}}} + 
\nonumber\\ 
\left.
   7.8336\,{{10}^{-10}}\,{x^{{\frac{84}{5}}}} - 
   4.56\,{{10}^{-11}}\,{x^{{\frac{96}{5}}}} + 
   2.66\,{{10}^{-12}}\,{x^{{\frac{108}{5}}}} - 
        1.68\,{{10}^{-13}}\,{x^{24}}+{{\rm O}(x)}^{{\frac{132}{5}}} 
\right].                 \label{PMC}
\eea
Its radius of convergence can be estimated to $\delta x^{12/5} = 14.3
\pm 0.4$, that is, $\delta x \simeq 3.0$.\footnote{A plot of this
  series sharply shows that it is very close to the TBA result for $x
  \lesssim 3$ and quickly departs from it for larger $x$.}  This
expansion (\ref{PMC}) can also be obtained by conformal perturbation
theory \cite{Zamo}. The series coefficients are then expressed in
terms of integrals of correlators of the perturbing conformal field,
that is, the thermal field of dimension $d_{\Phi}=4/5$ in the
three-state Potts model. The computation of these integrals is very
laborious, except for the first ones, which in some cases admit
explicit expressions as series \cite{KlasMel}. A related computational
method is the {\em truncated conformal-space approach}, in which one
only takes a finite dimensional subspace of the Hilbert space of
possible states, namely, a number of low-lying states, to numerically
diagonalize the Hamiltonian.  In the limit of increasing the number of
states, this approach is equivalent to a numerical evaluation of the
integrals giving the coefficients of the perturbation series
\cite{KlasMel}. We do not think worthwile to dwell in detailed
computational methods since the simple numerical integration of the
TBA equation (\ref{a2}) suffices to show the monotonicity of $\C$ and
$\tilde\C$.

\section{Discussion}

For a $2d$ field theory, one can introduce two monotonic dimensionless
functions, namely, $\C$ and $\tilde\C$, derived from the $2d$ relative
entropy and the $1d$ quantum entropy, respectively. It has been shown
that $\C$ is universal when the couplings are strongly relevant, that
is, with dimension $1 < y \leq 2$.  They include the thermal
perturbations of the unitary minimal models of conformal field theory
(CFT), except for the Ising model, which we have also studied,
notwithstanding. In contrast, $\tilde\C$ is always universal, since it
only depends on the universal finite-size correction to the free
energy.  Given that general theorems may not be particularly useful if
the quantities that they involve cannot be computed in practise,
considerable time and effort has been devoted to compute $\C$ and
$\tilde\C$ for a variety of models.  In consequence, we have been able
to show for them that those functions are monotonic.

The dimensionless entropies $\C$ and $\tilde\C$ play a similar r\^ole
to Zamolodchikov's $c$ function, constraining the structure of the RG
flow, but they have a clear physical origin, unlike Zamolodchikov's
$c$ function.  The existence of a monotonic function is usually argued
on the grounds of the irreversible nature of the RG flow, which in the
coarse-grained formulation implies a loss of information on
microscopic degrees of freedom \cite{ZamoC,Cardy}.  This idea inspired the
adaptation of Boltzmann's $H$ theorem to the RG flow in our previous
work \cite{I-OC}.  It has been shown here that this philosophy gives
rise to the entropic functions $\C$ and $\tilde\C$, which are
computable for a wide range of models. The non-perturbative
computation of Zamolodchikov's $c$ function is much harder and in fact
it does not seem to have been carried out for any fully interacting
model.  For all these reasons, the entropic $C$ functions proposed
here arguably provide a new perspective in the long-standing problem
of the irreversibility of the RG.  Nevertheless, in comparison with
Zamolodchikov's $c$ theorem, it must be remarked that universality of
$\C$, the entropic function more similar to Zamolodchikov's $c$
function, has been proved only for deformations of the critical theory
by fields with dimension $0\leq d_\Phi <1$ (strongly relevant), while
Zamolodchikov's theorem covers the entire range of $d_\Phi$.

Our entropic monotonicity theorem for the dimensionless relative entropy is 
\begin{equation}
x{d {\cal C}\over d x} = 
{\beta^2\over y}\int\!\! d^2z\,\langle :\!\Theta(z)\!: \, 
:\!\Theta(0)\!: \rangle.
\end{equation}
Even though it resembles Zamolodchikov's $c$ theorem it is not quite
the same: The correlator of $\Theta$'s in the second term appears
integrated.  Furthermore, a detailed calculation of Zamolodchikov's
function $c(m)$ for the free boson or fermion shows that they differ
from the respective values of ${\cal C}(m) = {\cal C}(x)|_{\beta=1}$.
The essential discrepancy actually has a geometrical origin: A crucial
step in the proof of Zamolodchikov's theorem relies on the assumption
of rotation symmetry \cite{Zamo}, which does not exist on the
cylinder. Therefore, the theorem does not hold on it.  However, the
absence of rotation symmetry is traded for the appearance of a new
parameter, the width $\beta$, which replaces the distance to the
origin in Zamolodchikov's theorem and is used in the derivation of the
entropic monotonicity theorems.

The reason for the introduction of a finite geometry is to have an IR 
scale to define a dimensionless relative entropy. We have used the cylinder 
because of its thermodynamic interpretation. Of course, other finite 
geometries are possible. For example, one can use a sphere. Its radius is 
then the IR scale. The advantage is that rotation symmetry is preserved 
on the sphere and Zamolodchikov's theorem holds. With this new geometry 
the monotonicity theorem for the dimensionless relative entropy 
would still involve an integral of 
the correlator of $\Theta$'s but a relation of $\C$ with Zamolodchikov's 
$c$ function seems more feasible. At least in conformal perturbation theory 
one should be able to perform that integral in terms of the IR scale and 
a direct comparison with Zamolodchikov's theorem could be possible. 

The existence of several monotonic functions prompts the question of
which one is preferable.  It is intuitively clear that a unique
definition of monotonic function is not possible: The RG itself is not
unique and one can choose a variety of RG parameters. Correspondingly,
if $\C(x)$ is monotonic a monotonic change of the independent variable
$x$ will transform it into a different monotonic function. We might
then consider what happens at the boundary, $x=0$ or $x \rightarrow
\infty$.  The point $x=0$ is the RG fixed point and it is sensible to
define a function related to it, as are the dimensionless relative
entropy $\C$ or Zamolodchikov's $c$ function. The dimensionless
absolute entropy $\tilde\C(x)$ is defined irrespective of the fixed
point and actually diverges there.  We could as well demand good
behaviour in the limit $x \rightarrow \infty$.  This condition is
fulfilled by $\tilde\C$ but may not be fulfilled by $\C$, owing to the
bulk term.  It is quite possible that a minor modification of the
definition of $\C$ may remove the bulk term and make it well behaved
in the limit $x \rightarrow \infty$ as well as at $x=0$.

\section*{Acknowledgments}

I acknowledge hospitality at the Nuclear Research Institute of Dubna
(Russia), where this work was started, and partial support under Grant
PB96-0887.  I thank D.~O'Con\-nor for conversations in the early stages
of this work, M.A.R.~Osorio, M.~Laucelli Meana and J.~Puente
Pe\~nalba for conversations on finite temperature QFT, A.~Bystko and
A.~Fring for conversations and correspondence on the TBA, and  A.~Petkou 
and A.~Dom\'{\i}nguez for latter conversations.

\appendix

\section{Calculation of finite-size corrections with 
the Euler-MacLaurin formula}

For free models the energy series can be evaluated with the
Euler-MacLaurin summation formula,\footnote{This is a common method to
  convert sums to integrals. However, since the function summed
  $\epsilon(n)$ diverges when $n \rightarrow \infty$ a preliminary
  regularization is required. A convenient form is to sum up to some
  arbitrary number $N \gg 1$, which for a chain can be the number of
  sites.  This UV regularization renders meaningful the formal
  manipulations that follow. However, we do not need to be definite on
  the UV regularization for our focus is on universal quantities.}
\begin{equation}
\sum_{n=0}^{\infty}\epsilon(n) =
\int\limits_{0}^{\infty}{dn}\,\epsilon(n) + 
{1\over 2}\,\epsilon(0) - {1\over 12}\,\epsilon'(0) + 
{1\over 720}\,\epsilon'''(0) - {1\over 30240}\,\epsilon^{(v)}(0) + \cdots,
\label{EM}
\end{equation}
where $\epsilon(n)$ are single-particle energies. For the Gaussian model 
the first term 
can be proved to be proportional to $L$ with the change of variable 
$p=2\pi\,n/L$ and leads to 
the integral in Eq.~(\ref{e0}). One can see that all the
odd derivatives of $\epsilon(n)$ vanish at $n=0$ because it is 
an even function. It is natural, because 
the finite size corrections are exponentially negligible 
when $L \rightarrow \infty$ and hence non analytic. Since
every derivative pulls out a power of
$2\pi/L$, the subsequent series of powers of $1/L$ must have 
vanishing coefficients.

In the scaling zone $m\,L \ll 1$ 
the finite-size corrections are not negligible, despite the previous 
argument.
To evaluate these corrections 
we can nevertheless make use of the Euler-MacLaurin expansion but 
using a non-zero value for the point at which the derivatives are computed, 
in the following form, 
\begin{equation}
\sum_{n=1}^{\infty}\epsilon(n) =
\int\limits_{1}^{\infty}{dn}\,\epsilon(n) + 
{1\over 2}\,\epsilon(1) - {1\over 12}\,\epsilon'(1) + 
{1\over 720}\,\epsilon'''(1) - {1\over 30240}\,\epsilon^{(v)}(1) + \cdots.
\label{EM1}
\end{equation}
Now the series can be cast as an expansion in powers of $m\,L$.
The reason why this trick works can be understood in several ways. 
One is that the derivatives $\epsilon^{(2k+1)}(n)$ as functions of $m$
are ill behaved for small $n$. They converge to the null function for $n=0$ 
but non uniformly. It is actually safer to choose the argument $n$ 
of the derivatives larger than in Eq.~(\ref{EM1}), $n=3$ or 4 say. Then 
the Euler-MacLaurin expansion converges very fast and the terms displayed 
above suffice to match the coefficients in Eq.~(\ref{EG}) with about 
ten decimal places.

Alternatively, one may focus on the fact that for $m=0$ the function
$\epsilon(n)\propto |n|$ is singular at $n=0$; its derivatives
eventually diverging there. This is naturally an IR divergence, which
does not exist for $m \neq 0$. However, one must be careful to
evaluate $\epsilon(n)$ at $n \neq 0$ before taking $m=0$, or in other
words, one must introduce an IR cutoff and evaluate the
Euler-MacLaurin expansion at that point.  Of course, the result shall
be inpependent of the precise value of the cutoff, although its
convergence properties are greatly affected by it.  Within the realm
of classical mathematics, it is interesting to recall that Legendre
met a similar problem when he attempted to evaluate elliptic integrals
numerically with the Euler-MacLaurin expansion. Since the integrand is
an even function of the integration variable at the limits 0 and
$\pi/2$, the odd derivatives vanish and the Euler-MacLaurin formula
implies that the elliptic integral is equal to any of its rectangular
approximations. The paradox was solved by Poisson, who showed that in
this case the remainder term does not tend to zero as the number of
terms increases and hence the series does not converge. If we further
consider that the bulk free energy of the Gaussian or Ising models on
a finite chain can be expressed as elliptic integrals, we may
appreciate that Legendre actually encountered an IR divergence without
being aware of the need of regularization.

For the Ising model (\ref{E0I}) the odd derivatives of $\epsilon(n)$ 
do not vanish 
at $n=0$ but it also is necessary to choose $n=3$ or 4 for fast convergence. 

\section{Calculation of $\langle T_{ab} \rangle$ 
on the cylinder for free models}

For free field theories the expectation values of the components of the 
stress tensor can be expressed in terms of the Green function. Thus
for a bosonic field
\be
\langle \Theta \rangle = m^2\,\langle \vf^2 \rangle = 
m^2\,\lim_{z \rightarrow 0}G_\beta(z,\bar{z}).
\ee
We use complex notation, $z = x_1+i\,x_2$.
The Green function on a cylinder, $G_\beta(z,\bar{z})$ is non trivial. 
Its Fourier transform 
includes a sum over discrete momenta in the compact direction,
\be
G_\beta(z,\bar{z}) = {1\over\beta} \sum_{n=-\infty}^{\infty} 
\int\limits_{-\infty}^{\infty} \frac{dk}{2\pi}\,
\frac{e^{i\,(\omega_n\,x_1+k\,x_2)}}{\omega_n^2+k^2 + m^2},
\label{FourG}
\ee
where the allowed frequencies for bosons are 
$\omega_n = \frac{2\pi}{\beta}\,n$. 
It can be transformed into a more manageable form by the use of 
the proper-time 
representation \cite{Br}. In this representation the integral over $k$ is 
elementary and one is left with the sum over $n$ and the integral 
over proper time. After performing a convenient Poisson resummation one 
obtains
\be
G_\beta(z,\bar{z}) = \sum_{n=-\infty}^{\infty} 
\int\limits_{0}^{\infty} \frac{ds}{4\pi s}\,
e^{-m^2 s-\frac{|z-n\,\beta|^2}{4 s}}.      \label{Gz}
\ee
Since the Green function on the plane is just 
$$G_\infty(z,\bar{z}) = \frac{1}{2\pi}\,K_0(m|z|),$$
which is the 
$n=0$ term in the sum (\ref{Gz}), this sum can be interpreted as the
solution of the field equation for a point source by the method of
images.  It can be expressed in terms of the Jacobi theta function
$$
\theta_3(\nu,\tau) = \sum_{n=-\infty}^{\infty} u^n\,q^{\frac{n^2}{2}},
\quad u = e^{2\pi i\nu}, \quad q = e^{2\pi i\tau},
$$
as
\be
G_\beta(z,\bar{z}) = \int\limits_{0}^{\infty} \frac{ds}{4\pi s}\,
e^{-m^2\,s-\frac{|z|^2}{4 s}}\,
\theta_3\left(-\frac{i\,x_1\,\beta}{4\pi s},\frac{i\,\beta^2}{4\pi s}\right).
\ee
Then the Poisson resummation realizes the duality property of $\theta_3$.

The formal ${z \rightarrow 0}$ limit of $G_\beta$ is easily 
taken,
\be
G_\beta(0)= \lim_{z \rightarrow 0}G_\beta(z,\bar{z}) = 
\sum_{n=-\infty}^{\infty} 
\int\limits_{0}^{\infty} 
\frac{ds}{4\pi s}\,e^{-m^2 s-\frac{(n\,\beta)^2}{4 s}}.
\ee
It contains the logarithmic divergence 
$$
G_\infty(0)= \frac{1}{2\pi}\,K_0(0) = 
\int\limits_{0}^{\infty} \frac{ds}{4\pi s}\,e^{-m^2 s}.
$$
Taking into account the integral representation of modified Bessel functions
\be
K_{\nu}(z) = \frac{1}{2}\,\left(\frac{z}{2}\right)^{\nu} 
\int\limits_{0}^{\infty}
\frac{ds}{s}\, s^{-\nu}\,e^{-s-\frac{z^2}{4 s}},
\ee
one obtains
\be
G_\beta(0)= \frac{1}{2\pi} \sum_{n=-\infty}^{\infty} K_0(n\,m\,\beta).
\ee

The computation of $\langle T \rangle$ requires a little more work, for 
\be
\langle T \rangle = -4\,\langle ({\partial_z}\vf)^2 \rangle = 
4\,\lim_{z \rightarrow 0}\partial_z^2G(z).
\ee
{}From (\ref{Gz}),
\be
\partial_z^2G_\beta(z,\bar{z}) = 
\sum_{n=-\infty}^{\infty} \int\limits_{0}^{\infty} 
\frac{ds}{4\pi s}\left(\frac{\bar{z}+n\,\beta}{4\,s}\right)^2 \,
e^{-m^2 s-\frac{|z-n\,\beta|^2}{4 s}}.
\ee
Hence,
\be
\partial_z^2G(0)= \lim_{z \rightarrow 0}\partial_z^2G(z) =
\sum_{n=-\infty}^{\infty} \int\limits_{0}^{\infty} 
\frac{ds}{4\pi s}\,\frac{(n\,\beta)^2}{16\,s^2} \,
e^{-m^2 s-\frac{(n\,\beta)^2}{4 s}} =
-\frac{m^2}{8\pi} \sum_{n=-\infty}^{\infty} K_2(n\,m\,\beta).
\ee
It contains a quadratic divergence in $K_2(0)$.

For Majorana fermions one could start from the known expressions of their 
stress tensor but it is simpler to 
consider them as bosons with antiperiodic boundary conditions and 
use again the Fourier transform (\ref{FourG}) with allowed frequencies 
$\omega_n = \frac{2\pi}{\beta}\left(n+\frac{1}{2}\right)$. 
Before Poisson resummation $G_\beta$ can be expressed in terms of 
$$
\theta_2(\nu,\tau) = \sum_{n=-\infty}^{\infty} u^{n+\frac{1}{2}}\,
q^{\frac{1}{2}{\left(n+\frac{1}{2}\right)^2}}.
$$
The Poisson resummation transforms $\theta_2$ into its dual $\theta_4$, 
$$
\theta_4(\nu,\tau) = \sum_{n=-\infty}^{\infty}(-)^n\, 
u^n\,q^{\frac{n^2}{2}},
$$
which is like the bosonic $\theta_3$ but with an additional alternating sign. 

We can write the final result in a condensed notation,
\bea
\langle \Theta \rangle =  
\pm\frac{m^2}{2\pi} \sum_{n=-\infty}^{\infty} (\pm)^n\,K_0(n\,m\,\beta),\\
\langle T \rangle =  
\pm\frac{m^2}{2\pi} \sum_{n=-\infty}^{\infty} (\pm)^n\,K_2(n\,m\,\beta),
\eea
where the upper signs stand for bosons and the lower signs for fermions. 

However, to have well defined expressions we must further 
introduce a regularization 
that removes the divergences in $K_0(0)$ and $K_2(0)$. It is customary 
to begin defining normal-ordered composite fields, namely, 
\bea
:\!\Theta\!:\; = m^2 :\!\vf^2\!:,   \\
:\!T\!:\;  = -4 :\!({\partial_z}\vf)^2\!:,
\eea
in the sense of a point splitting regularization and a substraction of 
the divergent part, computed with the Wick prescription. It amounts to 
the substraction of $G_\infty(0)=\pm{1}/({2\pi})\,K_0(0)$ or 
$4\,{\partial_z^2}G_\infty(0)=\pm{m^2}/({2\pi})\,K_2(0)$. Point splitting 
on a lattice yields 
\bea
K_0(0) = -\left(\ln\frac{m\,a}{2}+\gamma\right) + {\rm O}(a^2),\\
K_2(0) = \frac{2}{m^2\,a^2} - \frac{1}{2} + {\rm O}(a^2).
\eea
Considering $a \sim 1/\Lambda$ we have, for example, that according to 
Eq.~(\ref{FT}) 
\be
e_0 = \pm{m^2\over 4\pi}\left[K_0(0) - K_2(0)\right] \sim 
\pm{1\over 4\pi} \left\{-2\,\Lambda^2 + 
{m^2} \ln {2\,\Lambda\over m} 
+ {m^2} \left(\gamma - \frac{1}{2}\right) + {\rm O}(\Lambda^{-2})\right\}.
\ee

\end{document}